\documentclass[conference]{IEEEtran}
\IEEEoverridecommandlockouts

\usepackage{cite}
\usepackage{amsmath,amssymb,amsfonts}
\usepackage{algorithmic}
\usepackage{graphicx}
\usepackage{textcomp}
\usepackage[dvipsnames,table]{xcolor}
\usepackage{booktabs}
\usepackage{url}
\def\BibTeX{{\rm B\kern-.05em{\sc i\kern-.025em b}\kern-.08em
    T\kern-.1667em\lower.7ex\hbox{E}\kern-.125emX}}

\usepackage{mathpartir, semantic, bussproofs}
\usepackage[ruled,vlined,linesnumbered]{algorithm2e}
\usepackage{subcaption}
\usepackage{tikz}
\usepackage{tabularx}
\usepackage{makecell}
\usepackage[hidelinks]{hyperref}
\usetikzlibrary{arrows.meta, positioning, shapes.misc}

\usepackage[dvipsnames]{xcolor}
\usepackage{listings}

\definecolor{codebg}{RGB}{248,248,248}     % very light gray
\definecolor{codeframe}{gray}{0.75}
\definecolor{codekw}{RGB}{0,0,150}        % dark blue
\definecolor{codeid}{RGB}{50,50,50}       % identifiers
\definecolor{codecm}{gray}{0.4}           % comments
\definecolor{codestr}{RGB}{0,120,0}       % strings
\definecolor{codenum}{RGB}{180,0,0}       % numbers

\lstdefinestyle{code}{
  backgroundcolor=\color{codebg},
  basicstyle=\ttfamily\footnotesize\color{codeid},
  keywordstyle=\color{codekw}\bfseries,
  commentstyle=\itshape\color{codecm},
  stringstyle=\color{codestr},
  numberstyle=\tiny\color{codecm},
  numbers=left,
  stepnumber=1,
  numbersep=8pt,
  frame=single,
  rulecolor=\color{codeframe},
  columns=fullflexible,
  keepspaces=true,
  tabsize=2,
  showstringspaces=false,
  breaklines=true,
  xleftmargin=0.75em,
  xrightmargin=0.75em,
  captionpos=b
}

\lstdefinelanguage{CUDA}{
  language=C++,
  morekeywords={
    __global__, __device__, __host__, __shared__, __constant__,
    blockIdx, threadIdx, blockDim, gridDim, warpSize
  }
}

\newcommand{\ignore}[1]{}

\definecolor{reviseblue}{RGB}{0,0,0}        % dark blue (HTML #00008B)
\newcommand{\revise}[1]{{\color{reviseblue}#1}}

\newcommand{\typeword}[1]{\mathbf{#1}}

\newcommand{\fps}{\typeword{FP32}}

\newcommand{\is}{\typeword{Int32}}

\newcommand{\name}{CuLifter}
\newcommand{\sirname}{SSIR}

\newcommand{\typerecov}{type recovery}

\lstdefinelanguage{CUDA}{
  language=C++,
  morekeywords={__global__, __device__, __shared__, __syncthreads, warpSize}
}

\newenvironment{keeplisting}%
  {\par\addvspace{\medskipamount}\noindent
   \begin{minipage}{\linewidth}\lstset{aboveskip=0pt,belowskip=0pt}}%
  {\end{minipage}\par\addvspace{\medskipamount}}

\begin{document}

\title{\name{}: Lifting GPU Binaries to Typed IR}
%\subtitle{\normalsize{MICRO 2026 Submission
%    \textbf{\#2617} -- Confidential Draft -- Do NOT Distribute!!}}

%% Authors — uncomment and fill in for camera-ready
%\author{...}

\author{
\IEEEauthorblockN{Jisheng Zhao\textsuperscript{*}}
\IEEEauthorblockA{\textit{Georgia Institute of Technology}\\
%\textit{Georgia Institute of Technology}\\
Atlanta, USA \\
jishengz@gmail.com}

\and

\IEEEauthorblockN{Huanzhi Pu\textsuperscript{*}} 
\IEEEauthorblockA{\textit{Georgia Institute of Technology}\\
%\textit{Georgia Institute of Technology}\\
Atlanta, USA \\
hpu8@gatech.edu}

\and

\IEEEauthorblockN{Shinnung Jeong} 
\IEEEauthorblockA{\textit{Georgia Institute of Technology} \\
%\textit{Georgia Institute of Technology}\\
Atlanta, USA \\
sjeong306@gatech.edu}

% \and
% alignment fix
\and[\hfill\break\hspace*{\fill}]

\IEEEauthorblockN{Chihyo Ahn} 
\IEEEauthorblockA{\textit{Georgia Institute of Technology} \\
%\textit{Georgia Institute of Technology}\\
Atlanta, USA \\
ahnch@gatech.edu}

\and

\IEEEauthorblockN{Hyesoon Kim} 
\IEEEauthorblockA{\textit{Georgia Institute of Technology} \\
%\textit{Georgia Institute of Technology}\\
Atlanta, USA \\
hyesoon@cc.gatech.edu}

\thanks{\textsuperscript{*}Jisheng Zhao and Huanzhi Pu contributed equally to this work.}
}

\vspace{-10pt}

\maketitle

\begin{abstract}
GPU compilers merge all data types into a single unified register
file, erasing the type information that binary-analysis tools rely
on. We show that type recovery from this untyped register file is
the central challenge of GPU binary lifting. We present \name{},
a SASS-to-LLVM IR lifting framework that recovers register types
via constraint propagation with conflict detection, reconstructs
explicit control flow, and aggregates multi-instruction patterns.
Across eight benchmark suites spanning
open-source applications, vendor libraries, and optimized ML
runtimes, \name{} successfully lifts \revise{all} 11{,}977
kernels to valid LLVM IR. Among the testable set, we
achieve more than 90\% execution correctness, verified via the CPU
backend. An ablation 
study confirms that type recovery is the only step required to produce 
compilable IR: disabling it causes 86.9\% of kernels to execute incorrectly.

\end{abstract}

\begin{IEEEkeywords}
GPU binary lifting, LLVM IR, SASS, type recovery, program analysis
\end{IEEEkeywords}

%\footnotetext[1]{These authors contributed equally.}

%%
%% Paper body — all sections in body.tex
%%
% body.tex — single-file paper body for \name{}: Sections I–IX
% Tool name placeholder: \name{} (defined in main.tex)
% IR name placeholder:   \sirname{} (defined in main.tex)
% Data placeholders:     \TODO{...} (defined in main.tex)

%% ─────────────────────────────────────────────────────────────────────────
\section{Introduction}

NVIDIA GPUs dominate high-performance and parallel computing,
and CUDA~\cite{cuda} has become the de facto standard for GPU
programming across ML frameworks (e.g.,
PyTorch~\cite{pytorch}, vLLM~\cite{vllm}) and HPC
components~\cite{gpu_survey:2014}. Performance-critical CUDA
kernels increasingly ship as pre-compiled SASS binaries without
source code, including vendor libraries (cuDNN, TensorRT, cuBLAS,
NCCL), optimized ML runtimes, and closed-source HPC components.

Binary lifting translates machine code back into a higher-level IR
amenable to compiler analysis. For CPU binaries, lifted IR enables
a wide range of applications: security analysts use it to detect
vulnerabilities in closed-source
software~\cite{cpu_binary:2012:sp}, malware researchers apply
taint analysis and symbolic execution on lifted
code~\cite{cpu_binary:2011:bap}, and systems researchers
retarget legacy binaries to new
architectures~\cite{cpu_binary:2000:uqbt} or harden them with
control-flow integrity~\cite{cpu_binary:2016:asplos}.
These applications are supported by a mature tool ecosystem:
static lifters such as McSema~\cite{mcsema} and
rev.ng~\cite{revng} recover LLVM IR from x86, dynamic lifters
such as BinRec~\cite{binrec} capture execution traces for
recompilation, and a recent systematization of knowledge evaluates
these tools across downstream
applications~\cite{cpu_binary:2020:sp}.

No comparable infrastructure exists for GPU binaries.
NVIDIA provides disassemblers (\texttt{nvdisasm},
\texttt{cuObjDump}), but these output only flat, untyped assembly
without control-flow or type information.
NVBit~\cite{nvbit} and SASSI~\cite{sassi} instrument
SASS dynamically and at compile time, respectively, but neither
lifts the binary to an analyzable IR.
CuAsmRL~\cite{cuasmrl:cgo2025} optimizes SASS instruction
schedules via reinforcement learning, and
CASS~\cite{arxiv:25:CASS} transpiles SASS to AMD RDNA3 assembly using
LLMs, achieving 37.5\% assembly-level translation accuracy.
Both operate on SASS but neither produces a typed, retargetable
IR suitable for compiler analysis. As a result, the analyses
routinely applied to CPU binaries (vulnerability detection,
performance modeling, code portability) remain unavailable for
the GPU binaries that dominate modern computing workloads.

A key obstacle to GPU binary lifting is \emph{type recovery}:
determining the data type (e.g., \texttt{Float32},
\texttt{Int32}, pointer) that each register holds at each
program point. Type recovery is a well-known challenge for CPU
binaries as well, where integers and pointers share the same
general-purpose registers and memory aliasing obscures data
types. A long line of work addresses this problem, progressing
from constraint-based
inference~\cite{tie,secondwrite:pldi2013} to lattice-based
propagation~\cite{retypd}, probabilistic
prediction~\cite{osprey}, and behavior-capturing
types~\cite{trex:2025:usenix}. However, CPU ISAs still provide
partial structural cues: separate register files for
floating-point and vector values, explicit transfer instructions
(\texttt{movd}, \texttt{cvtsi2sd}), and calling conventions that
constrain parameter types. GPU ISAs provide far fewer cues.
SASS stores all values in a single unified register file, and
the same register can be consumed as \texttt{Float32} by one
instruction and as \texttt{Int32} by the next, with no
intervening transfer. Without recovering these implicit type
boundaries, a lifter cannot emit IR at all: every LLVM IR
instruction requires explicit type annotations, so unresolved
types block code generation entirely.

Consider a compiled \texttt{rsqrt()} call:
\begin{keeplisting}
\begin{lstlisting}[caption={Compiled \texttt{rsqrt()}: register
\texttt{R25} changes type with no transfer instruction.},
                   label={lst:intro-rsqrt}]
FADD     R25, R25, 1e-8          // Float32: R25 += eps
MUFU.RSQ R24, R25                // Float32: rsqrt(R25)
IADD3    R0, R25, -0xd000000, RZ // Int32: exponent bits
ISETP.GT P0, PT, R0, 0x727fffff  // Int32: range check
@!P0 BRA 0x340                   // branch out of range
FMUL     R0, R25, R24            // Float32: R0=R25*R24
\end{lstlisting}
\end{keeplisting}
\vspace{-2mm}

Register \texttt{R25} is used as \texttt{Float32} in lines~1--2
and~6, but reinterpreted as \texttt{Int32} in line~3, with no
transfer instruction between them. A lifter must detect this
implicit type boundary and insert an explicit \texttt{bitcast}. Otherwise, it cannot assign a LLVM type to \texttt{R25}.

Our key insight is that GPU binary lifting is a
\emph{structure-recovery} problem. GPU compilation erases three
kinds of structure (control flow, instruction-level semantics,
and data types), a framing consistent with recent formalizations
of binary lifting~\cite{fible:2025:oopsla}. Among these, type
recovery is the most critical and the focus of this paper.

We present \name{}, a framework that recovers analyzable LLVM IR
from NVIDIA SASS. We target LLVM IR because its mature ecosystem
of analysis and optimization passes applies directly to the
recovered code, and its backends enable retargeting to other
architectures (e.g., x86 CPUs). To recover each lost structure,
\name{} applies a targeted technique: it infers register types
by propagating type constraints from typed opcodes along def-use
edges, reconstructs explicit control flow from SASS's predicated
execution model, and reassembles multi-instruction patterns
(e.g., carry chains, reciprocal approximations) into single
high-level operations.

We make the following contributions:
\begin{enumerate}
    \item We identify \textbf{type recovery} as the central
    challenge of GPU binary lifting and solve it via constraint
    propagation with conflict detection.
    \item We adapt $\psi$-function-based SSA
    construction~\cite{stoutchinin2001psi} to SASS, extending it
    for SIMT-specific predication and convergence barriers.
    \item We build \name{}, a complete SASS-to-LLVM IR lifter
    supporting SM75, SM90, SM100, and SM120, and evaluate it on
    11{,}977 SASS kernels across eight benchmark suites. \name{}
    successfully lifts all kernels to valid LLVM IR and obtains
    more than 90\% end-to-end execution correctness in every
    suite.
\end{enumerate}

%% ─────────────────────────────────────────────────────────────────────────
\section{Background}
\label{sec:back}

\subsection{SASS as a Low-Level Representation}
\label{sec:SASS}

\begin{table}[htbp]
\caption{Key structural differences between SASS and LLVM}
\footnotesize
\begin{center}
\begin{tabular}{l|l|l}
    & \textit{SASS} & \textit{LLVM IR}\\%\hline\hline
    \midrule

operand & physical register & virtual SSA value \\
%\hline
type system & untyped (unified & strongly typed \\
            & register file)   & (i32, i64, float, ptr, \ldots) \\
%\hline
control flow & predicated execution & explicit conditional \\
             & + unconditional BRA  & branch on SSA value \\
%\hline
memory access & arch-specific ops  & generic load/store \\
              & (LDG, LDS, LDC, \ldots) & with address space specified \\
%\hline
addressing & reg + imm offset & typed pointer \\ %(GEP) \\
%\hline
tensor op & HMMA, IMMA, & target-specific \\
          & HGMMA (SM90+) & intrinsics \\
%\hline
tensor precision & encoded in opcode & abstracted in \\
                 & suffix (modifiers)          & intrinsic signature \\
\end{tabular}
\label{tab:saasvsllvmir}
\vspace{-3mm}
\end{center}
\end{table}

SASS~\cite{nvsass} is the native machine code for NVIDIA GPUs. During compilation, CUDA source code is first compiled to PTX~\cite{nvptx}, a hardware-independent virtual assembly that preserves structured control flow and type information, and then lowered to SASS, a hardware-specific assembly where much of this structure is lost. SASS is optimized for specific GPU architectures (e.g., Turing SM75, Ampere SM80, Hopper SM90), with instruction encoding, register files, memory layout, and synchronization differing across generations. 
Table~\ref{tab:saasvsllvmir} summarizes the key structural differences between SASS and LLVM IR.

LLVM IR~\cite{llvm} is a strongly typed, platform-independent assembly in static single assignment (SSA) form~\cite{ssa}. It bridges high-level code and machine-specific assembly, with support for x86, ARM, RISC-V, NVPTX, and others.

\subsection{SIMT Execution Model}
\label{sec:simt}
The NVIDIA GPU architecture is built upon the Single Instruction,
Multiple Threads (SIMT) execution model. Threads are grouped
into warps of 32 threads that execute in lockstep. When threads
within a warp encounter a data-dependent branch, divergence
occurs: the warp executes each branch path sequentially, while
predicate registers mask off inactive threads. As a result,
SASS encodes most conditional behavior through predicated
execution rather than explicit branch instructions. Starting
with the Volta architecture, NVIDIA introduced Independent
Thread Scheduling, which gives each thread its own program
counter and call stack, allowing threads to make independent
forward progress even after divergence. However, to regain
lockstep efficiency once threads reconverge, NVIDIA employs
explicit convergence instructions. On Volta and later
architectures, BSSY and BSYNC are used. On pre-Volta architectures, SSY and SYNC serve this purpose.
%The NVIDIA GPU architecture is based on the Single Instruction, Multiple Threads (SIMT) execution model. Threads are grouped into warps of 32 that execute in lockstep. When threads diverge at a data-dependent branch, the warp executes each path sequentially, masking inactive threads via predicate registers. \revise{Since Volta architecture, NVIDIA introduced Independent Thread Scheduling, granting each thread its own program counter and call stack to enable independent forward progress.} To regain lockstep efficiency after divergence, convergence instructions (\texttt{BSSY}/\texttt{BSYNC} on Volta and later, \texttt{SSY}/\texttt{SYNC} on earlier architectures) are used to mark reconvergence points.

\subsection{Predicated SSA and $\psi$-Functions}
\label{sec:psifunctions}

Standard SSA construction~\cite{ssa} assumes every instruction in a basic block executes unconditionally. Predicated code violates this: two instructions that write the same register under complementary predicates both ``execute'' in the same block, and standard SSA rename kills the first definition, producing wrong code.

Stoutchinin and de Ferri\`{e}re~\cite{stoutchinin2001psi} introduced the $\psi$-function (\emph{psi-function}) for IA-64 predicated code: where a $\phi$-node selects a value based on which control-flow edge was taken, a $\psi$-function selects based on a predicate register's value. A predicated write \texttt{@P0 R0 = x} followed by \texttt{@!P0 R0 = y} becomes $\texttt{R0} = \psi(\texttt{P0}: x,\; \texttt{!P0}: y)$, semantically equivalent to a \texttt{select} instruction in LLVM IR. This technique is well-studied in the IA-64 compilation literature~\cite{warter1993reverse, carter1999pssa, chuang2003phi}. We apply the same principle to GPU predicated code in Section~\ref{sec:ssir}.
%% ─────────────────────────────────────────────────────────────────────────
\section{Challenges in GPU Binary Lifting}
\label{sec:hard}

CPU binary lifters rely on separate register files, explicit
branch instructions, and publicly documented
ISAs~\cite{cpu_binary:2016:asplos, cpu_binary:2011:bap,
cpu_binary:2019:survey}. SASS provides none of these: its
unified register file carries no type information, predication
replaces branches for short conditionals, and the ISA is
proprietary and changes across GPU generations. We characterize each
challenge below, starting with type recovery.

\subsection{Type Evidence Is Sparse and Must Be Propagated}

\label{sec:hard:types}
%\begin{table}[t]
%\caption{Register file organization across ISAs. CPU architectures encode type in register names; SASS does not.}
%\footnotesize
%\begin{center}
%\begin{tabular}{l|l|l|l}
%\textbf{ISA} & \textbf{Int regs} & \textbf{FP regs} & \textbf{Type signal} \\\hline\hline
%x86-64 & RAX--R15 & XMM0--15 & Free \\
%AArch64 & X0--X30 & V0--V31 & Free \\
%IA-64 & r0--r127 & f0--f127 & Free \\
%\textbf{SASS} & \multicolumn{2}{l|}{\textbf{R0--R255 (all types)}} & \textbf{Must recover} \\
%\end{tabular}
%\label{tab:regfiles}
%\end{center}
%\end{table}

Type recovery is the key technical challenge of GPU binary
lifting. On CPUs, separate integer and floating-point register
files provide type information for free. On SASS, a unified
register file holds \texttt{Int32}, \texttt{Float32},
\texttt{Float64} (as register pairs), pointers, packed
\texttt{Float16$\times$2}, and boolean predicates interchangeably.
The register name carries no type information. Because LLVM IR
requires explicit types on every instruction, a lifter that
cannot distinguish \texttt{Float32} from \texttt{Int32} cannot
emit valid IR at all.

\ignore{
This is the key technical challenge of GPU binary lifting. % and does not arise in CPU binary lifting. 
%As shown in Table~\ref{tab:regfiles}, 
Every major CPU ISA uses separate register files for integers and floating-point. For example, x86-64, AArch64, and IA-64 each have separate integer register files (RAX--R15, X0--X30, and r0--r127, respectively) and floating-point register files (XMM0--15, V0--V31, and f0--f127, respectively). Therefore, CPU binary lifters never asks \textit{``is this value float or integer?''} --- the register file answers for free. However, on SASS, unified register file hold \texttt{Int32}, \texttt{Float32}, \texttt{Float64} (as register pairs), pointers, packed \texttt{Float16$\times$2}, and boolean predicates interchangeably. The register name carries no type information.
}

Most SASS instructions encode type constraints in their opcodes
(e.g., \texttt{FADD} operates on floats, \texttt{ISETP}
produces a boolean predicate), providing direct type evidence
for ${\sim}$90\% of all instructions
(Section~\ref{sec:eval:typechar}). However, two challenges
remain. First, the remaining \emph{type-transparent}
instructions (moves, bitwise logic, and memory operations) pass
values without revealing their type, requiring propagation along
def-use chains to resolve them. Second, GPU compilation
introduces \emph{intentional type conflicts}: deliberate
reinterpretation of float bits as integers for
optimization~\cite{lomont2003}. We detect over 1.5 million
such conflicts across our benchmarks
(Section~\ref{sec:eval:typechar}). Incorrect type assignment
produces silent numerical corruption: a 32-bit float and an
integer with the same bit pattern differ by orders of magnitude
but cause no runtime fault.

As shown in Listing~\ref{lst:intro-rsqrt}, register
\texttt{R25} is defined by \texttt{FADD} (Float32), consumed
by \texttt{MUFU.RSQ} (Float32), then read by \texttt{IADD3}
(Int32), and finally used again by \texttt{FMUL} (Float32). A
forward-only approach forces Float32 onto \texttt{IADD3},
producing wrong integer arithmetic. The correct solution
detects the type conflict
($\{\fps\} \cap \{\is\} = \emptyset$) and inserts a
\texttt{bitcast} at the boundary.

\textbf{Key intuition:} GPU type recovery is a propagation
problem with conflict detection. The density of type seeds and
the length of transparent chains determine how far evidence
must travel, and empty intersections reveal implicit type
boundaries. We formalize this in
Section~\ref{sec:typepropagation} and characterize propagation
reach empirically in Section~\ref{sec:eval:typechar}.

\subsection{Control Structure Is Not Explicit}
\label{sec:hard:control}
Control recovery in SASS is fundamentally challenging due to a structural gap in the representation of control flow. LLVM IR expects an explicit control structure in which conditional execution is implemented strictly via branch instructions. In contrast, SASS uses an implicit model in which almost any instruction can be predicated. Instead of a conditional branch instruction, SASS models conditional branching via predicate-guarded execution of an unconditional \texttt{BRA}. The actual branch conditions are contained implicitly within the predicate register, whose set-predicate instruction could be many instructions away. In addition, this implicit nature of predication further complicates SSA construction, because a predicated definition updates a value only for the threads whose predicate is true, while the remaining threads implicitly preserve the old value, breaking standard data-flow tracking assumptions.

\textbf{Key intuition:} control flow in SASS is heavily carried by predicate flow. To bridge this structural gap and produce a valid SSA form, the lifter must handle the predication semantics and reconstruct the missing explicit logic from predication correctly.

\subsection{Semantic Operations Are Collapsed into
Multi-Instruction Patterns}
\label{sec:hard:patterns}

GPU compilers collapse high-level operations into
multi-instruction patterns whose structure changes across GPU
generations. A 64-bit subtraction threads a borrow predicate
across two \texttt{IADD3}/\texttt{IADD3.X} instructions. If a lifter handles each instruction in isolation, the carry chain
breaks and the wrong result is produced. On SM90, a single
\texttt{cp.async} in CUDA becomes an orchestrated sequence:
\texttt{FENCE.\allowbreak VIEW.\allowbreak ASYNC.\allowbreak S} issues a memory fence,
\texttt{LDGSTS.\allowbreak E.\allowbreak 128} performs a fused global-to-shared copy,
and \texttt{LDGDEPBAR}/\texttt{DEPBAR}/\texttt{BAR.\allowbreak SYNC} form an async completion chain. A lifter that treats these
independently would decompose \texttt{LDGSTS} into a
synchronous load and store and discard the barrier chain as
no-ops. SM90 also introduces packed-conversion instructions
such as \texttt{F2FP.\allowbreak PACK\allowbreak\_AB\allowbreak\_MERGE\allowbreak\_C}, which converts two
\texttt{Float32} values to Float8, saturates, packs both into the upper 16 bits, and merges with a third operand's lower 16
bits, all in a single opcode. No direct LLVM intrinsic exists. The lifter must synthesize the full decomposition.
\textbf{Key intuition:} Without pattern recognition, lifted IR
is correct in behavior but opaque in intent. We present the
full pattern catalog in Section~\ref{sec:archspecific}.

%GPU compilers lower high-level operations into multi-instruction patterns that vary across generations. For example, on SM52, a 32-bit multiply-add uses a three-instruction \texttt{XMAD} chain, whereas SM75 expresses it as a single \texttt{IMAD}. A 64-bit subtraction threads a borrow predicate across two \texttt{IADD3}/\texttt{IADD3.X} instructions; if a lifter handles each instruction in isolation, the carry chain breaks and the wrong result is produced. Reciprocal approximation patterns combine \texttt{MUFU.RCP} with integer-side bias corrections that cross type boundaries.

%\textbf{Key intuition:} Without pattern recognition, lifted IR is correct in behavior but opaque in intent. We present the full pattern catalog in Section~\ref{sec:archspecific}.

\subsection{Implicit Register Dependencies}
\label{sec:hard:implicit}

%SASS instructions hide register operands. 
%First, Each operand names only
%the first register of a consecutive group; the remaining registers
%are implicit, determined by the opcode. For example,
%\texttt{LDG.E.64 R4, [R2]} names two registers but actually
%accesses four: it defines \texttt{R5:R4} and reads \texttt{R3:R2}.
%A lifter that sees only \texttt{R4} and \texttt{R2} misses half the
%data dependencies.

%This matters for correctness: if another instruction writes
%\texttt{R5} between the load and its consumer, the lifter must
%detect the conflict --- but it cannot, because \texttt{R5} never
%appears in the disassembly. Getting this wrong silently corrupts
%values rather than producing an error.

%\HK{This actually can be a great example of challegnes of GPU. there is a bug report for this code.--> cannot add this due to the space }

SASS instructions
may access registers that are not explicitly named in the
disassembly. In many cases, an operand denotes only the first
register of a contiguous register group, while the remaining
registers are determined implicitly by the opcode semantics. For
example, \texttt{LDG.E.64 R4, [R2]} appears with only two
explicit registers, but it actually defines \texttt{R5:R4} and
reads \texttt{R3:R2}. A conventional lifter that models only
the explicitly named operands therefore misses half the data
dependencies. This directly affects correctness. If another instruction writes \texttt{R5} between the load and its consumer, the lifter must
detect the conflict, but it cannot, because \texttt{R5} never
appears in the disassembly. Getting this wrong silently corrupts
values rather than producing an error. Tensor-core instructions demonstrate an extreme case. On
SM75, \texttt{HMMA.16816.F32} names four registers but actually
accesses fourteen. On SM90,
\texttt{HGMMA.64x128x16.F32} accesses 164 registers through
four named operands.

\textbf{Key intuition:} On CPU ISAs, every accessed register is
named in the instruction encoding. On SASS, the lifter must
reconstruct hidden registers from per-opcode tables before
building data dependencies. Otherwise SSA, def-use analysis, and type propagation all operate on incomplete information.

%Tensor-core instructions are the extreme case.
%Listing~\ref{lst:hmma-implicit} shows \texttt{HMMA.16816.F32},
%which names four registers in the disassembly but actually reads and
%writes fourteen. On SM90, \texttt{HGMMA.64x128x16.F32} accesses
%196~registers through four named operands. Table~\ref{tab:implicit-regs}
%summarizes the scale of hidden registers across instruction classes.

%\begin{lstlisting}[caption={HMMA.16816.F32 names 4 registers but
%accesses 14. R8--R11 are both read (as C accumulator) and written
%(as D result); R3--R5 and R13 are read but never named.},
%                   label={lst:hmma-implicit}]
%// What the disassembly shows:
%HMMA.16816.F32  R8, R2, R12, R8

%// What the hardware accesses:
%//  writes: R8, R9, R10, R11       (D: 4 result registers)
%//  reads:  R2, R3, R4, R5         (A: 4 input registers)
%//          R12, R13                (B: 2 input registers)
%//          R8, R9, R10, R11       (C: 4 accumulator registers)
%\end{lstlisting}

% \begin{table}[t]
% \caption{Implicit register dependencies. ``Named'' = registers visible
% in the disassembly; ``Accessed'' = registers actually read or written.}
% \footnotesize
% \centering
% \begin{tabular}{l|c|c|c}
% \textbf{Instruction} & \textbf{Named} & \textbf{Accessed} & \textbf{Implicit} \\\hline\hline
% \texttt{LDG.E R4, [R2]}       & 2 & 3  & 1 \\
% \texttt{LDG.E.128 R4, [R2]}   & 2 & 6  & 4 \\
% \texttt{DFMA R4, R2, R6, R8}  & 4 & 8  & 4 \\
% \texttt{HMMA.16816.F32}        & 4 & 14 & 10 \\
% \texttt{HGMMA.64x128x16.F32}  & 4 & 196 & 192 \\
% \end{tabular}
% \label{tab:implicit-regs}
% \vspace{-3mm}
% \end{table}

\subsection{Device Functions Are Embedded} %, Not Separate}
\label{sec:hard:devicefunc}

In CPU binaries, every callable function is a distinct ELF
symbol with its own entry point, and lifters discover function
boundaries directly from the symbol table. On SASS, the
compiler embeds device functions \emph{inside} the kernel
function. They are not inlined (they retain their own control
flow with \texttt{CALL} and \texttt{RET} instructions), but
they share the same SASS section and have no separate symbol
table entry.

\begin{keeplisting}
\begin{lstlisting}[caption={Device function embedded inside the kernel function.
The call target \texttt{0x780} is not a separate ELF symbol; it is
a basic block inside the same function's instruction stream.},
                   label={lst:device-func}]
// Kernel body
0x210: CALL.ABS.NOINC 0x780  // call device function
0x220: FADD    R8, R8, R6    // return continues here
       ...                   // (kernel body continues)
// Device function (same SASS section, no ELF symbol)
0x780: LDG.E   R2, [R4]      // entry point
0x790: FFMA    R2, R2, R5, R6
0x7a0: RET                   // returns to 0x220
\end{lstlisting}
\end{keeplisting}

\noindent
After CFG construction, the kernel body and the device function
appear as \emph{disconnected subgraphs} within a single SASS
function. \texttt{CALL} does not create a normal CFG edge. It saves the return address to a register pair and branches, but
does not duplicate the device function.
%A lifter that ignores this structure sees the device function
%blocks as unreachable dead code and misses register definitions
%that flow back through \texttt{RET} to the call site.

\textbf{Key intuition:} Embedded device functions must be
recovered as call/return structure, not treated as unreachable
CFG fragments. Otherwise the lifter loses inter-block definitions that flow back through \texttt{RET}.
%% ─────────────────────────────────────────────────────────────────────────
\section{Solution Approach}
\label{sec:requirements}

This section presents our solutions to the two key technical problems at an algorithmic level, independent of SASS-specific details. Section~\ref{sec:design} describes the GPU-specific implementation.

\subsection{Type Recovery via Constraint Propagation}
\label{sec:typepropagation}

We formulate type recovery as a constraint-propagation problem over type sets. Each value $v$ carries a set of candidate types $\mathcal{T}(v) \subseteq L$, where $L = \{\textit{Bool}, \textit{Int32}, \textit{Float32}, \textit{Int64}, \textit{Float64}, \textit{Float16}, \textit{Int128}\}$. All values start unconstrained: $\mathcal{T}(v) = L$.

\textbf{Phase 1: Detect type seed.} Instructions with known
type semantics (e.g., floating-point add $\Rightarrow$
\textit{Float32}, integer compare $\Rightarrow$ \textit{Bool})
narrow their operands' sets. Most SASS instructions are
type-seeding (Section~\ref{sec:hard:types}), but the remaining
type-transparent instructions create gaps that propagation must
bridge.

\textbf{Phase 2: Propagate type information.} Type-transparent
instructions (Section~\ref{sec:hard:types}) propagate
constraints by intersecting the candidate sets of related
operands along def-use edges. A worklist-based fixpoint
iterates until no set changes.

\textbf{Phase 3: Detect conflicts and insert bitcasts.} When
intersection produces $\emptyset$, the algorithm has discovered
an \emph{implicit type boundary}, a point where the compiler
reinterprets bits across types with no explicit instruction.
The conflict resolution rule is: (1)~The value retains its
\textbf{definition-site type}. (2)~The conflicting
\textbf{use site} receives an explicit \texttt{bitcast}.
(3)~Once marked conflicted, subsequent type assignments
\textbf{overwrite} rather than intersect, preventing cascading
false conflicts.

For example, Figure~\ref{fig:type-walkthrough} traces the recovery phase through Listing~\ref{lst:intro-rsqrt}. A register holding a float value is read as an integer by a subsequent instruction. Seeding establishes the definition-site type as \textit{Float32}. When the integer use requests \textit{Int32}, the intersection $\{\textit{Float32}\} \cap \{\textit{Int32}\} = \emptyset$ triggers a conflict: the value keeps \textit{Float32}, and the integer use receives a \texttt{bitcast fp32 to i32}. A later float use of the same register requires no bitcast because the definition-site type is already correct.

\begin{figure}[t]
\centering
\scriptsize
\begin{tabular}{@{}l@{\quad}c@{\quad}l@{}}
\textbf{Instruction} & \textbf{Type set after seed} & \textbf{Emitted IR} \\\hline
float add & $\{$Float32$\}$ & \texttt{\%v = fadd float \%v, 1e-8} \\
float rsqrt & $\{$Float32$\}$ & \texttt{\%u = call float @rsqrt(\%v)} \\
\rowcolor{red!8}
int subtract & $\{$Int32$\}$ \enspace $\cap\{$Float32$\}=\emptyset$ & \texttt{\%vi = \textbf{bitcast} float \%v to i32} \\
& \textit{conflict!} & \texttt{\%r = sub i32 \%vi, 0xd000000} \\
float multiply & $\{$Float32$\}$ \enspace (no conflict) & \texttt{\%w = fmul float \%v, \%u} \\
\end{tabular}
\caption{Type recovery walkthrough for Listing~\ref{lst:intro-rsqrt}. The float-add definition establishes \textit{Float32}. The integer use triggers an empty intersection. A \texttt{bitcast} is inserted at the use site. The subsequent float use requires no bitcast.}\label{fig:type-walkthrough}
\vspace{-5mm}

\end{figure}

After convergence, values with multiple remaining candidates are resolved by priority ($\textit{Int32}$ preferred), a conservative choice that preserves bit patterns for a subsequent \texttt{bitcast}. The key quantity is \emph{propagation reach}: how many transparent hops evidence must travel. We measure this across real GPU kernels in Section~\ref{sec:eval:typechar}.

\subsection{Predicated SSA via Select Insertion}
\label{sec:predsolution}

We apply the $\psi$-function technique (Section~\ref{sec:psifunctions}) to GPU predicated code: each predicated write is isolated into a synthetic basic block and merged via a \texttt{select}:

\begin{keeplisting}
\begin{lstlisting}[caption={Predicated if-else: two writes to the same register under complementary predicates, and the corrected SSA with select insertion.},
                   label={lst:pred-ssa}]
// Before (flat predicated code):
p  = (x > 0)
r  = x << 1            // unconditional write
@!p r = x + 100        // conditional write
use(r)                 // which r reaches use?

// After (select insertion):
p  = (x > 0)
r1 = x << 1
r2 = x + 100
r  = select(p, r1, r2)
use(r)
\end{lstlisting}
\end{keeplisting}

\noindent
This is semantically equivalent to a $\psi$-function: $r = \psi(p\!: r_1,\; \neg p\!: r_2)$. On GPUs, three additional SIMT-specific constructs require handling beyond IA-64: (1)~predicated thread termination must become a conditional return rather than function exit, (2)~dual-predicate branches must separate the branch condition from the convergence annotation, and (3)~hardware convergence barriers must be treated as metadata. We detail these in Section~\ref{sec:ssir}.

%% ─────────────────────────────────────────────────────────────────────────
\section{Recovering Analyzable Structure in \name{}}
\label{sec:design}

%This section describes how \name{} recovers the three kinds of structure identified in Section~\ref{sec:hard}: 

This section describes three structure recovery techniques identified in Section~\ref{sec:hard}. %: explicit control flow (Section~\ref{sec:ssir}), normalized semantic operations (Section~\ref{sec:archspecific}), and typed value flow (Section~\ref{sec:formaltype}).

\subsection{Recovery Ordering}
\label{sec:recoveryprocedure}

\begin{figure*}[t]
    \centering
\includegraphics[width=6in]{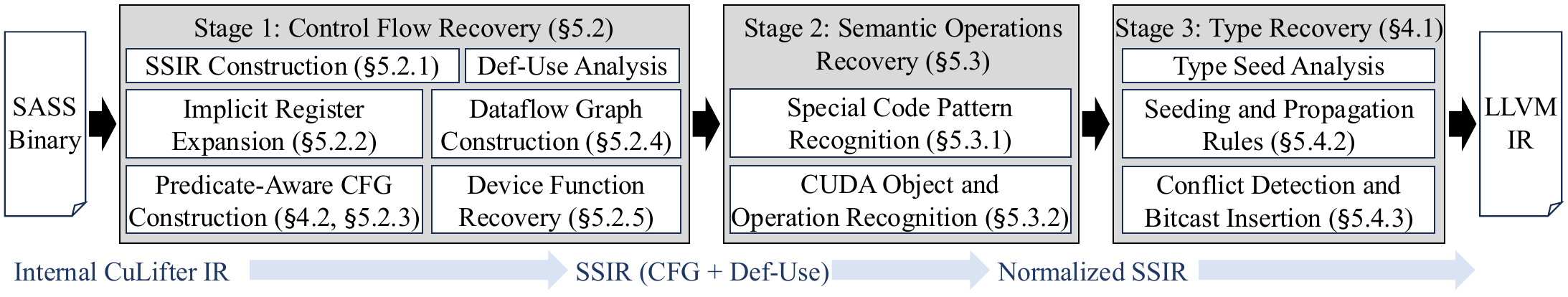}
    \caption{The \name{} lifting pipeline. SASS is transformed into LLVM IR through three stages:
(1)~Control flow recovery with SSA conversion and dataflow analysis. (2)~Semantic operation recovery with architecture-specific normalization (multiply idioms, reciprocal chains, copy propagation, dead code). (3)~Type recovery, and LLVM IR emission.}
%Pass names correspond to implementation classes.}
    \label{fig:culifter}
\end{figure*}

The two techniques from Section~\ref{sec:requirements},
together with pattern normalization, must execute in a specific
order as shown in Figure~\ref{fig:culifter}: control-flow
reconstruction first (to build the CFG with SSA form), semantic
operation recovery second (to normalize and to aggregate
multi-instruction patterns), and type recovery last (to
propagate over the normalized graph). Control-flow
reconstruction must come first because both pattern matching
and type propagation traverse def-use edges, which require a
complete CFG in SSA form. Semantic operation recovery must
precede type analysis because the type lattice operates over
aggregated values. Typing a carry-chain pair before aggregation would mistype the borrow predicate as an independent integer
rather than as internal to a 64-bit operation. Stages~1 and~2
are architecture-dependent because instruction encodings,
control codes, and predicate semantics differ across GPU
generations.

%\begin{algorithm}
%\footnotesize
%\caption{RecoverStructure: instruction stream to typed IR}
%\label{alg:recover}
%\SetKwFunction{BuildCFG}{BuildCFG}
%\SetKwFunction{LiftToSSA}{LiftToSSA}
%\SetKwFunction{NormalizePatterns}{NormalizePatterns}
%\SetKwFunction{SeedTypes}{SeedTypes}
%\SetKwFunction{PropagateTypes}{PropagateTypes}
%\SetKwFunction{EmitIR}{EmitIR}

%\KwIn{$S$: flat instruction stream}
%\KwOut{$M$: typed IR module}

%\tcp{Phase 1: CFG + predicated SSA (\S\ref{sec:predsolution})}
%$G \leftarrow$ \BuildCFG{$S$} \tcp*{identify branches, barriers, predicate roles}
%\LiftToSSA{$G$} \tcp*{split predicated blocks, insert selects, SSA rename}

%\tcp{Phase 2: pattern normalization}
%\NormalizePatterns{$G$} \tcp*{aggregate multi-instruction idioms}

%\tcp{Phase 3: type recovery (\S\ref{sec:typepropagation})}
%$T \leftarrow$ \SeedTypes{$G$} \tcp*{seed from typed instructions}
%\PropagateTypes{$G$, $T$} \tcp*{fixpoint: intersect, detect conflicts, insert bitcasts}
%$M \leftarrow$ \EmitIR{$G$} \\
%\Return{$M$}
%\end{algorithm}

\subsection{Control Structure Recovery}
\label{sec:ssir}

\subsubsection{\sirname{} Construction}

\sirname{} (SASS Static IR) is the IR that bridges the
architecture-specific and architecture-independent stages.
\sirname{} represents six levels: function metadata, basic
blocks, instructions with opcodes and operands, operand type
information, CUDA-specific objects (cooperative groups, warps),
and synchronization operations. During parsing, \sirname{}
preserves architecture-specific information including nested
opcodes (e.g., \texttt{XMAD.MRG}), operand properties (e.g.,
\texttt{R0.reuse}), and control codes. Instruction
normalization expands nested opcodes to 3-address form. One SASS instruction may produce several \sirname{} instructions.

Several architecture-specific issues are resolved at this
stage. Branch targets are realigned from control-code lines to
valid instruction addresses. Predicate registers are treated
as normal operands, and predicated instructions are isolated
into single-instruction basic blocks for downstream SSA
construction. Compiler-inserted infinite loops after
\texttt{EXIT} are removed. NVCC places these as safety padding to prevent threads from falling through to invalid
instructions, but they are not reachable control flow. On
Volta and later architectures, instructions with multiple
definitions (carry-out predicates) are split into separate
instructions.

\subsubsection{Implicit Register Expansion}
\label{sec:implicit-expansion}

\name{} expands every instruction to its full set of accessed registers using per-opcode tables. Each register group is treated as a single SSIR operand with sub‑register access. This representation preserves the original type constraints. This expansion is applied in the parsing phase. 
%We resolve hidden register dependencies at parse time, before any analysis pass runs. During \sirname{} construction, each instruction's operands are expanded to include all implicit registers based on per-opcode tables.

For 64-bit operations, the named register is replaced with an
explicit register pair: \texttt{R4} becomes \texttt{R5:R4}. For
128-bit operations (\texttt{LDG.128}, \texttt{STS.128}), a
register quad \texttt{R7:R6:R5:R4} is created. For tensor-core
instructions, the expansion is shape-dependent:
\begin{itemize}
    \item \texttt{HMMA.16816.F32} (SM75): the shape
    \texttt{m16n8k16} determines D=4, A=4, B=2, C=4 registers.
    \item \texttt{HGMMA.64x128x16.F32} (SM90): the warp-group
    shape \texttt{m64} with \texttt{N=128} determines
    D=64, A=4, B=32, C=64 registers.
\end{itemize}

A register group is treated as a single \sirname{} operand with
sub-register access, so that type analysis can preserve the type
constraint (\texttt{R3:R2} as an int64 address operand in
\texttt{LDG}) while assigning types to individual registers
(e.g., \texttt{R3} is later used as int32 by \texttt{IADD3}).

\subsubsection{Predicate-Aware CFG Construction}
\label{sec:cfg}

We identify basic-block leaders using standard rules (function
entry, branch targets, fall-through successors) plus one
GPU-specific rule: whenever the predicate guard changes, a new
block starts. This isolates predicated regions for later
analysis.

On SASS, conditional control flow is encoded as
predicate-guarded execution of unconditional instructions, so
the lifter must recover the implicit condition. Unconditional
branches become \texttt{Br} terminators. For predicated
\texttt{BRA}/\texttt{EXIT} patterns, \name{} extracts the
predicate guard and constructs a \texttt{CondBr} with both
taken and fall-through successors. For example, a predicated
\texttt{@P0 EXIT} terminates only threads satisfying the
condition. The remaining threads fall through to the next instruction. A non-SIMT-aware lifter treats this as
unconditional termination, making the fall-through unreachable.
SIMT-specific instructions (\texttt{BSSY}, \texttt{BSYNC},
\texttt{WARPSYNC}) are kept as ordinary instructions with
metadata for later stages.

%Unconditional branches become \texttt{Br} terminators; predicated \texttt{BRA}/ \texttt{EXIT} patterns become \texttt{CondBr} with both taken and fall-through successors. SIMT-specific instructions (\texttt{BSSY}, \texttt{BSYNC}, \texttt{WARPSYNC}) are kept as ordinary instructions with metadata for later stages.

%\begin{lstlisting}[caption={Warp divergence encoded via predicated EXIT.},
%                   label={lst:divergence-example}]
%ISETP.GE.AND    P0, PT, R0, c[0x0][0x158], PT
%@P0 EXIT
%SHL             R6, R0.reuse, 0x2
%\end{lstlisting}

%To illustrate, consider a predicated \texttt{@P0 EXIT}.
%\texttt{EXIT} terminates only threads satisfying the condition;
%the rest fall through to the next instruction. A non-SIMT-aware
%lifter treats this as unconditional termination, making the
%fall-through unreachable. To preserve control-flow information,
%\name{} reconstructs the divergent branch with both successors.

A more complex case arises with dual-predicate branches, where
two independent predicates control the same instruction:

\begin{keeplisting}
\begin{lstlisting}[caption={Dual-predicate branch: execution guard and branch condition are separate predicates.},
                   label={lst:dual-pred-branch}]
ISETP.EQ.AND P1, PT, R24, 0x1, PT // P1 = (R24 == 1)
ISETP.NE.AND P0, PT, R2, RZ, PT   // P0 = (R2 != 0)
@!P0 BRA P1, 0x2810               // !P0: branch on P1
STL          [R1+0x4], RZ         // fall-through
\end{lstlisting}
\end{keeplisting}

\texttt{@!P0 BRA P1} has two predicate roles: \texttt{@!P0}
is the SIMT execution guard, and \texttt{P1} is the branch
condition. We represent this as a nested predicated branch. Treating either field as primary produces an incorrect CFG.

\subsubsection{Dataflow Graph Construction}
\label{sec:dfg}

On top of the CFG, we build an intra-procedural def-use graph. Nodes are \sirname{} instructions. Edges represent register value flow from definitions to uses. The analysis includes predicate registers and special registers (e.g., \texttt{SR\_TID}) so that later passes can reason about SIMT-specific data flow. Because SSA guarantees each register has exactly one definition, the analysis is a simple two-pass scan: the first pass maps each register to its defining instruction. The second links each use to its definition in constant time.
\subsubsection{Device Function Recovery}
\label{sec:devicefunc-recovery}

After CFG construction and before SSA, we recover embedded
device functions by identifying disconnected subgraphs
unreachable from the entry block. Call targets are resolved
through the ELF relocation table. Recovered device functions
are either extracted into separate \sirname{} functions or
inlined into the caller.

\ignore{
We recover embedded device functions after CFG construction, before
SSA:

\begin{enumerate}
    \item \textbf{Discover:}  BFS from the entry block identifies
    the kernel's reachable blocks.  Any remaining blocks form
    disconnected subgraphs. Each is a candidate device function.
    \item \textbf{Resolve call targets:}
  \texttt{CALL} instructions do not encode their target
    address directly. We read the ELF relocation table to    resolve each call site to its target address.

    \item \textbf{Extract or inline:}  Device functions called
    more than a threshold number of times ($> 16$) are extracted
    into separate \sirname{} \texttt{Function} objects with their
    own SSA scope.  Rarely-called device functions are inlined by
    cloning their blocks into the caller with remapped virtual
    addresses, restoring a single connected CFG.
\end{enumerate}
}
\noindent
Inlining is the common case: the CUDA compiler inlines most device
functions at the PTX level, so the few that survive to SASS are
typically small helpers. Extraction is reserved for functions with
many call sites (e.g., math library routines compiled as device
functions) to avoid code bloat.

\subsection{Semantic Operations Recovery}
\label{sec:archspecific}
%\subsection{Recovering Semantic Operations}

Stage~2 takes the \sirname{} from Stage~1 and (1)~aggregates instruction sequences implementing a logical operation, and (2)~recovers CUDA-level objects from synchronization and communication patterns.

\subsubsection{Special Code Pattern Recognition}
\label{sec:codepattern}

NVIDIA's compiler decomposes logical operations into
multi-instruction patterns, either for performance or because
64-bit values require 32-bit register pairs.  Lifting instruction-by-instruction produces verbose IR that loses the original intent.

\name{} uses a pattern matcher based on variable unification (shown in Algorithm~\ref{alg:graphmatch}). Each pattern template specifies expected opcodes and symbolic operand names. The matcher scans basic blocks, attempts to map template variables to actual registers (unify), and rewrites matched sequences into aggregated pseudo-instructions. Table~\ref{tab:specialpatterns} lists the code patterns in the matcher.
% Five dedicated normalizations (CarryChainAgg, XmadToImad, OpModTransform, SRSubstituteReverse, ReciprocalNorm)(shown in Table~\ref{tab:specialpatterns}) are always enabled.%; additional sequence-matching patterns for 64-bit arithmetic are defined but not yet enabled by default. 
% There are also architecture-specific issues for normalizations, i.e. the pattern template for different architectures are different due to high-level operations can be expressed as different instruction sequences across GPU generations.

\begin{algorithm}
\footnotesize
\caption{Variable-unification-based pattern identification}
\label{alg:graphmatch}
  \SetKwFunction{FindCandidates}{FindCandidates}
  \SetKwFunction{MatchOpcodes}{MatchOpcodes}
  \SetKwFunction{Unify}{Unify}

  \KwIn{\textit{BB}: basic block; \textit{PatternTable}: instruction-sequence templates}
  \KwOut{\textit{matches}: matched instruction groups}
    \textit{matches} $\leftarrow \emptyset$ \\
    \For{\textit{pattern} $\in$ \textit{PatternTable}} {
      \textit{candidates} $\leftarrow$ \FindCandidates{\textit{BB}, \textit{pattern.opcodes}} \\
      \For{\textit{candidate} $\in$ \textit{candidates}} {
        \textit{vars} $\leftarrow \emptyset$ \tcp*{variable bindings}
        \textit{ok} $\leftarrow$ \textbf{true} \\
        \For{(\textit{inst}, \textit{tmpl}) $\in$ zip(\textit{candidate}, \textit{pattern.templates})} {
          \If{$\neg$\MatchOpcodes{\textit{inst}, \textit{tmpl}}} {
            \textit{ok} $\leftarrow$ \textbf{false}; \textbf{break}
          }
          \If{$\neg$\Unify{\textit{inst.operands}, \textit{tmpl.vars}, \textit{vars}}} {
            \textit{ok} $\leftarrow$ \textbf{false}; \textbf{break}
          }
        }
        \If{\textit{ok}} {
          \textit{matches} $\leftarrow$ \textit{matches} $\cup$ \{(\textit{pattern.name}, \textit{candidate})\}
        }
      }
    }
    \Return{\textit{matches}}
\end{algorithm}

\begin{table}[htbp]
\caption{Normalizations implemented in \name{}.}
\footnotesize
\begin{center}
\renewcommand{\arraystretch}{1.1}
\begin{tabular}{l|p{5.2cm}}
\textbf{Pass / Pattern} & \textbf{Description} \\%\hline\hline
\midrule
\multicolumn{2}{l}{\textit{Code Pattern Normalization}} \\\hline
% CarryChainAgg & 32-bit carry-chain pair $\rightarrow$ 64-bit arthmatic operation aggregation \\ 
XmadToImad        & Rewrites the 3-instruction SM52 \texttt{XMAD} / \texttt{XMAD.MRG} / \texttt{XMAD.PSL.CBCC} idiom into a single \texttt{IMAD} \\
OpModTransform     & Expands the \texttt{.X4} address-scale modifier by inserting an explicit \texttt{SHL\,2} instruction \\
SRSubstitute & Scans constant-memory operands (\texttt{c[0][offset]}) for known \texttt{SR\_*} offsets (e.g.\ \texttt{SR\_TID.X} at \texttt{0x2C}); inserts explicit \texttt{S2R} instructions at first use and rewrites the operand to the fresh temp register \\
ReciprocalNorm     & Detects the \texttt{I2F}$\to$\texttt{MUFU}$\to$\texttt{IADD}/\texttt{IADD3}$\to$\texttt{F2I} magic-constant reciprocal chain emitted by NVCC; inserts \texttt{BITCAST} nodes around the integer-add step to preserve correct float/int reinterpretation semantics \\\hline
%\multicolumn{2}{l}{\textit{Sequence-matching patterns (defined, not yet enabled by default)}} \\\hline
%IADD3 + IADD3.X   & 32-bit carry-chain pair $\rightarrow$ 64-bit \texttt{IADD364} \\
%ISETP + ISETP.EX  & 64-bit integer compare across two predicate halves $\rightarrow$ \texttt{ISETP64} \\
%LEA + LEA.HI.X    & 64-bit effective-address computation $\rightarrow$ \texttt{LEA64} \\
%IMAD.WIDE         & Widening 32$\times$32$\rightarrow$64 multiply $\rightarrow$ \texttt{IMAD64} \\
%MOV + MOV         & Consecutive 32-bit moves of the same source $\rightarrow$ \texttt{MOV64} \\
%SHF (logical)     & \texttt{SHF.R} high-half extraction $\rightarrow$ \texttt{CAST64} or \texttt{SHR64} \\
%SHF + SHF / IMAD + SHF & 64-bit left-shift reconstruction $\rightarrow$ \texttt{SHL64} \\
%Pack64            & 32-bit register pair $\rightarrow$ 64-bit value via \texttt{PACK64} node \\
\multicolumn{2}{l}{\textit{Cooperative Group Object and Operations Recognition}} \\\hline
\textbf{Group creation} & Identified through thread mask manipulation instructions (e.g., \texttt{WARPSYNC}, \texttt{BAR.SYNC}) and predicate configurations that define participation masks.\\ 
\textbf{Sync operation} & \texttt{BAR.SYNC} with specific barrier IDs indicates thread-block synchronization; \texttt{WARPSYNC} with mask patterns identifies warp-level synchronization. \\ 
\textbf{Collective operations} & Cooperative reductions manifest as specific shuffle instruction sequences. \\ 
\textbf{Type differentiation} & Thread-block groups exhibit block-wide \texttt{BAR.SYNC} patterns; warp groups show \texttt{WARPSYNC~0xFFFFFFFF} and shuffle-based communication. \\
\end{tabular}
\label{tab:specialpatterns}
\end{center}
\end{table}

\subsubsection{CUDA Object and Operation Recognition}
\label{sec:cudaobjpattern}

CUDA-specific objects such as cooperative groups are not represented by explicit SASS opcodes; they emerge through specific synchronization and communication patterns. \name{} identifies them by scanning for characteristic opcode sequences (shown in Table~\ref{tab:specialpatterns}).

\subsection{Type Recovery}
\label{sec:formaltype}
%\subsection{Recovering Type Structure}
%\label{sec:type-recovery}

Stage~3 of CuLifter assigns architecture-independent types to all SSIR values using a constraint-propagation mechanism. 

\subsubsection{Type Domain and Lattice}

\[ L = \{\, \text{Bool},\; \text{Int32},\; \text{Float32},\; \text{Int64},\; \text{Float64},\; \text{Float16},\; \text{Int128}\; \} \]

To enable propagation through type-agnostic operations (such as register moves), we partition the lattice into width-based subsets (e.g., $N_{32} = \{\text{Int32}, \text{Float32}\}$, $N_{64} = \{\text{Int64}, \text{Float64}\}$). For each value $v$, we maintain a candidate type set $\mathcal{T}(v)$. Initially, $\mathcal{T}_0(v)$ is set to a top element representing all types. The lattice is extended with $\bot = \emptyset$ to denote a type conflict.

\begin{table}[t]
\footnotesize
\centering
\caption{Classification of SASS and their type signatures}
\label{tab:seed-class}
\begin{tabular}{l|l}
%\toprule
Category & Example Signature (Defs $\leftarrow$ Uses) \\
\midrule
Floating-point ALU & \texttt{FADD} $\;\{\text{Float32}\} \leftarrow \{\text{Float32}\}, \{\text{Float32}\}$ \\
Integer ALU & \texttt{IADD} $\;\{\text{Int32}\} \leftarrow \{\text{Int32}\}, \{\text{Int32}\}$ \\
Compare & \texttt{ISETP} $\;\{\text{Bool}, \text{Bool}\} \leftarrow \{\text{Int32}\}, \{\text{Int32}\}, \{\text{Bool}\}$ \\
Memory (Global) & \texttt{LDG} $\;\{\text{Num32}\} \leftarrow \{\text{Int64}\}$ \\
Memory (Shared) & \texttt{LDS} $\;\{\text{Num32}\} \leftarrow \{\text{Int32}\}$ \\
Type-transparent & \texttt{MOV, PACK64, SHR} $\; (\text{Propagate types only})$ \\
%\bottomrule
\end{tabular}
\end{table}

\subsubsection{Seeding and Propagation Rules}

Type evidence originates from two sources.

\paragraph{Seeding instructions.} Opcodes with fixed type semantics (e.g., \texttt{FADD} seeds \(\{\text{Float32}\}\), \texttt{ISETP} seeds \(\{\text{Bool}\}\)) narrow the candidate set of their defined value:
\[
\mathcal{T}(v) \leftarrow \mathcal{T}(v) \cap C(I),
\]
where \(C(I) \subseteq L\) is the type constraint of instruction \(I\). %Over all benchmarks, \(90.1\%\) of SASS instructions are type-seeding.

\paragraph{Type-transparent instructions.} Moves, bitwise logic, and memory operations propagate types without revealing them. For a transparent instruction with inputs \(x_1,\dots,x_k\) and output \(y\):
\[
\mathcal{T}(y) \leftarrow \bigcap_{i=1}^k \mathcal{T}(x_i), \qquad
\forall i,\; \mathcal{T}(x_i) \leftarrow \mathcal{T}(x_i) \cap \mathcal{T}(y).
\]

Table~\ref{tab:seed-class} categorizes SASS instructions according to their role.

% \begin{table}[!ht]
% \footnotesize
% \centering
% \caption{Classification of SASS instructions for type recovery}
% \label{tab:seed-class}
% \begin{tabular}{lll}
% \toprule
% Category & Example opcodes & Type constraint \\
% \midrule
% Floating-point ALU & \texttt{FADD, FMUL, FMA, MUFU.RSQ} & \(\{\text{Float32}\}\) or \(\{\text{Float64}\}\) \\
% Integer ALU & \texttt{IADD, IMAD, LOP3, ISETP} & \(\{\text{Int32}\}\) or \(\{\text{Bool}\}\) \\
% Compare & \texttt{ISETP, FSETP} & \(\{\text{Bool}\}\) \\
% Memory & \texttt{LDG, STG, LDS} & \(\{\text{ptr}\}\) (treated as \(\text{Int32}\)) \\
% Type-transparent & \texttt{MOV, AND, OR, XOR, SHL, } & none (propagate only) \\
%  & \texttt{SHR, PACK64} &  \\
% \bottomrule
% \end{tabular}
% \end{table}

\begin{figure}[t]
    \centering
\includegraphics[width=0.85\linewidth]{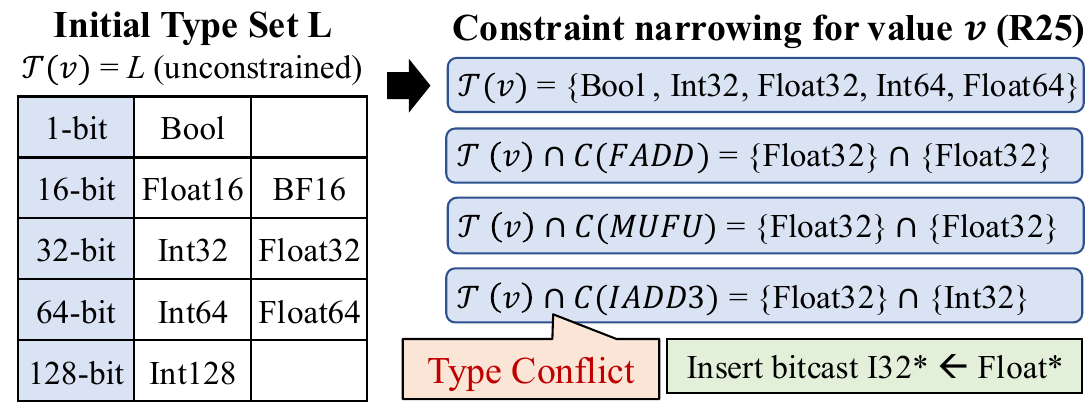}
    \caption{Iterative Propagation and Conflict Detection walkthrough for Listing~\ref{lst:intro-rsqrt}}
%Pass names correspond to implementation classes.}
    \label{fig:type_recovery}
\end{figure}

\subsubsection{Iterative Propagation and Conflict Detection}

The type recovery engine traverses the basic blocks of the control-flow graph, iteratively propagating constraints until it reaches a fixpoint or the iteration limit is reached. 

For type-transparent operations, types must be propagated between specific operands without narrowing to a fixed hardware type. Let $\mathcal{P}(I)$ be the set of propagation groups for an instruction $I$, where each group $G \in \mathcal{P}(I)$ contains operands that must share the same type.

During each iteration, the engine computes the intersection for every group:
\[
t_{\text{meet}} = \bigcap_{x \in G} \mathcal{T}(x)
\]
This updated constraint is then applied back to all operands in the group ($\forall x \in G : \mathcal{T}(x) \leftarrow t_{\text{meet}}$). 

A \emph{type conflict} occurs when propagation yields $t_{\text{meet}} = \emptyset$. This indicates an implicit type boundary, where the SASS compiler has reinterpreted the same hardware register as a different type without an explicit conversion instruction. To resolve the conflict and satisfy LLVM's strict type system, the lifter retains the definition-site type and makes the boundary explicit by inserting a \texttt{bitcast} operation at the conflicting use site. Figure~\ref{fig:type_recovery} gives an example that presents the type states for analyzing the code in listing 1.

\subsubsection{Resolving Ambiguous Types}

After fixpoint convergence, some values may still have $|\mathcal{T}(v)| > 1$ (multiple candidate types). CuLifter applies a fallback prioritizing integer types:
\[
\text{resolve}(v) = \begin{cases}
\text{Int32}, & \text{if } \text{Int32} \in \mathcal{T}(v), \\
\text{Int64}, & \text{if } \text{Int64} \in \mathcal{T}(v), \\
\text{Int128}, & \text{if } \text{Int128} \in \mathcal{T}(v), \\
t \in \mathcal{T}(v), & \text{otherwise.}
\end{cases}
\]

In the rare case, if the constraint set is empty because no type is propagated to it, the system defaults to Int32 to ensure compilable IR generation. 

% \subsubsection{Integration with SSIR}

% The type recovery pass operates on the SSIR (SASS Static IR) after control-flow reconstruction and pattern normalization (Section~\ref{sec:archspecific}). Each SASS register maps to an SSIR value; type-transparent instructions are identified via a precomputed table. The output is a typed LLVM IR where every value has a concrete type and implicit bitcasts are made explicit. The ablation study (Section~\ref{sec:eval:ablation}) confirms that type recovery is the only step required to produce compilable IR.

\subsection{Lowering to LLVM IR}
\label{sec:llvmirgen}

After type annotation and pattern normalization, we emit LLVM IR through per-opcode handlers. Table~\ref{tab:ssir2llvmir} shows the mapping: physical registers become typed SSA values, special registers become NVIDIA intrinsics, and memory-space qualifiers become LLVM address-space annotations.

\begin{table}[t]
\footnotesize
\centering
\caption{\sirname{} to LLVM IR mapping.}
%\begin{center}
\label{tab:ssir2llvmir}
\begin{tabular}{l|l}
\textbf{\sirname{}} & \textbf{LLVM IR} \\%\hline\hline
\midrule
Registers (R0, R1, ...)            & Logical operand with type \\
Special registers (\%tid, \%ctaid) & NV-specific intrinsics \\
Memory spaces (global, shared, local) & Explicit address spaces \\
Arithmetic/logic operators         & LLVM IR equivalents \\
Control flow (CondBra/Bra)         & CondBr/Br instructions \\
Barriers (BAR.SYNC)                & Sync intrinsics \\
\end{tabular}
%\end{center}
\end{table}

\name{} maps SASS tensor operations to LLVM representations (Table~\ref{tab:tensor-ops}). MMA instructions use NVIDIA's standard \texttt{llvm.nvvm.mma.*} intrinsics. Tensor memory operations lower to typed loads/stores with address-space annotations. Tensor reductions decompose into scalar equivalents. We support SASS tensor types \texttt{.F16}, \texttt{.F16x2}, \texttt{.TF32}, \texttt{.I8}, and \texttt{.I32}, mapped to LLVM vector types $\langle N \times \text{half} \rangle$, $\langle N \times \text{float} \rangle$, $\langle N \times \text{i8} \rangle$, and $\langle N \times \text{i32} \rangle$ respectively.
\begin{table}[!t]
\footnotesize
\centering
\caption{Formal mapping of tensor operations to LLVM}
\label{tab:tensor-ops}
\begin{tabular}{l|l}
%\toprule
\textbf{SASS operation} & \textbf{LLVM representation} \\
\midrule
MMA multiply-accumulate & \texttt{@llvm.nvvm.mma.*} \\
Tensor load & \texttt{load <type>, addrspace(N)} \\
Tensor store & \texttt{store <type>, addrspace(N)} \\
Tensor reduction (sum) & scalar FMA decomposition \\
Tensor reduction (max) & scalar compare/select \\
%\bottomrule
\end{tabular}
\end{table}

\subsection{Architecture Support}
\label{sec:archsupport}
We support SM75 (Turing), SM90 (Hopper), and SM100/SM120
(Blackwell), with legacy support for SM35 (Kepler) and SM52
(Maxwell).  Architecture support is localized to Stages~1 and~2: the parser handles different instruction encoding and predicate semantics, and the pattern library adds generation-specific idioms. Type analysis and LLVM generation are architecture independent. Adding a new generation requires extending the parser and pattern library only.

The main evolution across generations is the convergence mechanism. Pre-Volta architectures use reconvergence stacks (\texttt{SSY}/\texttt{SYNC}). Turing and later use barriers (\texttt{BSSY}/\texttt{BSYNC}). We handle both by abstracting convergence as opaque synchronization intrinsics.

\subsection{Type Accuracy}
In this section, we evaluate the accuracy of our type recovery directly. Our methodology is same as the evaluation for CPU binary type inference~\cite{tie, caballero2016typesurvey}. We compile with debug info (\texttt{nvcc -G -g}) and take DWARF types as ground truth. Since \texttt{-G} disables optimization and alters register allocation, we add a source-derived track that maps \texttt{.cu} kernel-signature types to constant-memory offsets via the CUDA ABI and compares against types inferred on optimized binaries. We report \emph{exact accuracy} (inferred type matches ground truth) and \emph{kind accuracy} (correct family---int, float, or bool---regardless of width).

We measured the type accuracies across HeCBench's 229 benchmarks. Across both SM75 and SM90, \name{} achieves 97.1\% kind accuracy and 87.4\% exact accuracy. We analyzed the instruction patterns leading to the drop in kind accuracy and exact accuracy and categorized them into four patterns. 

Type analysis rests on the assumption that each instruction is used strictly according to its semantics. As we show below, this is not always the case: the compiler often applies special computing patterns to optimize code and to improve hardware-unit utilization. Such cases manifest in two ways. In some, the mismatch surfaces as a \emph{type boundary}, which we discuss further in the type-boundary analysis (Section~\ref{sec:eval:typeboundary}). In others, it causes the type analysis to misidentify a register's type because of an incorrect assumption about how the instruction is used. 
% This misidentification is sometimes harmless, but in other cases it requires manual correction. In either situation, a data type may be misrepresented without violating semantic correctness.

\paragraph{Non-canonical Opcode Uses}
\label{sec:noncanonical}

In most cases, the compiler respects instruction semantics, using each opcode for its intended purpose on registers of the matching type. However, we observe that an instruction is sometimes used for a purpose that does not match its target data type, a choice the compiler makes as an optimization to improve utilization. This introduces errors and imprecision in the type analysis, because seed instructions are pinned to their corresponding data types. We touched on this in the motivation, where the motivating example (Listing~\ref{lst:intro-rsqrt}) shows an integer instruction operating on a floating-point value. Here we present the reverse scenario, in which a floating-point instruction is used to produce an integer value.

\begin{keeplisting}
\begin{lstlisting}[label={lst:fp-builds-int}]
// HFMA2 writes the integer value 4 into R5
HFMA2.MMA R5, -RZ, RZ, 0, 2.384185791015625e-07
// addr: R2 = base + R4*R5 = base + idx*sizeof(float)
// R5 reused as an i32 stride -> seed/use conflict
IMAD.WIDE R2, R4, R5, c[0x0][0x210]
\end{lstlisting}
\end{keeplisting}

In the Listing above, the seemingly arbitrary floating-point immediate encodes the integer value 4. The SASS-level semantics are misleading on their own and can be understood only by examining the surrounding instruction context. This contributes to both kind accuracy and exact accuracy. 

 The fix is simply to insert a reinterpretation cast at the \emph{type boundary}. The lifted IR faithfully reflects a type conflict that already exists in the input SASS, so correctness is not compromised.  We find that type boundaries carry rich information about the techniques the compiler employs, discussed in Section~\ref{sec:eval:typeboundary}.

\paragraph{Opaque pointer types}
\label{sec:unrecoverable}

At the binary level, a pointer is a register pair that holds a 64-bit address referencing another value. While type analysis can sometimes infer the pointee type, this information is not always available, which results in a drop in exact accuracy. 

\begin{keeplisting}
\begin{lstlisting}[label={lst:ptr-i64}]
// DWARF ground truth: float* A; (typed 64-bit ptr)
// R4: i32 index, R5: i32 stride -> R2:R3 = i64 address
IMAD.WIDE R2, R4, R5, c[0x0][0x210]
// R2:R3 packed into one i64 pointer (pointee type lost)
LDG.E     R0, [R2.64]
\end{lstlisting}
\end{keeplisting}

In this example, a pointer is read from constant memory and used directly in a memory load. We decide that the pointee type is unnecessary, and the pointer can be treated as a 64-bit integer.  We argue that low-level pointer arithmetic is greatly simplified by treating pointers as plain integers. This reduces exact accuracy, but the semantic correctness of the kernel is maintained.

\paragraph{Packed Datatypes}
To enable efficient data movement, the compiler often packs two 16-bit values into a single 32-bit register, and several instructions are designed to operate on these packed operands. For the bfloat type, for example, a common register layout is \texttt{bfloat16x2}, two bfloats packed together. Instructions such as \texttt{HADD2} add both halves simultaneously.
\begin{keeplisting}
\begin{lstlisting}[label={lst:packed-bf16}]
HADD2 R7, R7, R6
\end{lstlisting}
\end{keeplisting}

The instruction in the example performs a genuine packed-half add. To handle such cases, we assign the packed data type a scalar type of equivalent width, which is Float32 here. This simplifies the type analysis without breaking correctness, since HADD2 can unpack the data later in its IR implementation. Nevertheless, this choice leads to a drop in exact accuracy.

\paragraph{Wide Datatypes}
The compiler sometimes emits two instructions that operate on two 32-bit registers to perform a 64-bit operation. This pattern reduces the exact accuracy due to width mismatch.
\begin{keeplisting}
\begin{lstlisting}[label={lst:wide-i64}]
IMAD.MOV.U32 R12, RZ, RZ, c[0x0][0x180]
IMAD.MOV.U32 R13, RZ, RZ, c[0x0][0x184]
\end{lstlisting}
\end{keeplisting}
The instruction pattern loads a int64 at the semantic level, but at the low level it is emitted as two int32 loads. This gap can be narrowed by re-aggregating them into the original wider type. We currently do not re-aggregate those patterns because faithful lifting preserves the correctness of the lifted IR.
% \paragraph{Type Accuracy}
% As discussed above, our type analysis is challenged by patterns that can lead to incorrect or ambiguous type resolution. We therefore evaluate the accuracy of our type recovery directly. To evaluate type recovery independently of execution correctness, we adopt the standard methodology from CPU binary type inference~\cite{tie, caballero2016typesurvey}: we compile kernels with debug information (\texttt{nvcc -G -g}), extract DWARF types as ground truth, and compare them against the inferred types. Because \texttt{-G} disables optimizations and alters register allocation, we supplement this with a source-derived track that parses kernel signatures directly from \texttt{.cu} files, maps parameter types to constant-memory offsets via the CUDA ABI, and compares against types inferred on optimized binaries. We report five metrics: \emph{exact accuracy} (inferred type matches ground truth), \emph{kind accuracy} (correct type family---integer, float, or bool---regardless of width), \emph{conservativeness} (inferred type is a valid supertype), \emph{per-type-class accuracy}, and \emph{coverage} (fraction of ground-truth variables with an inferred type). 

%% ─────────────────────────────────────────────────────────────────────────

%% %% ─────────────────────────────────────────────────────────────────────────
%% CuLifter Section 6: Evaluation (Edited Version)
%% Based on version 7 PDF data. Changes marked with %%CHANGE comments.
%% ─────────────────────────────────────────────────────────────────────────

\section{Evaluation}
\label{sec:eval}

\subsection{Methodology}
\label{sec:eval:methodology}

We evaluate \name{} on eight benchmark suites totaling
11{,}977 kernels.  Table~\ref{tab:passrates} lists the evaluated suites and their kernel counts.
%reports the per-suite end-to-end execution-correctness results;
Table~\ref{tab:domain_distribution} shows the corresponding
breakdown by application domain, spanning Linear Algebra,
ML/DNN, Simulation, Science, Image/Vision, Graph, Crypto, and
Infrastructure workloads.  All results in this section are
based on SM90 (Hopper).\footnote{Due to limited GPU resources,
we focused evaluation on SM90.  SM75 results are qualitatively
similar.  We verified correctness on SM100/SM120 (Blackwell)
but omit those results for space.}

We compile the lifted LLVM IR to a target architecture (x86 or
Arm) and execute on the same inputs as the original binary.
Integer outputs are compared bitwise. Floating-point outputs use a relative tolerance of $10^{-5}$ for Float32, $10^{-2}$ for
Float16. A kernel passes if all outputs match.

\begin{table*}[t]
\caption{Evaluated benchmark suites. We evaluate \name{} on eight suites
totaling 11{,}977 kernels spanning ten application domains, ranging from
open-source applications to vendor libraries and optimized ML runtimes.}
\label{tab:passrates}
\footnotesize
\centering
\renewcommand{\arraystretch}{1.12}
\begin{tabularx}{\textwidth}{l|X|>{\raggedright\arraybackslash}p{4.8cm}|r}
\textbf{Suite} & \textbf{Desc} & \textbf{Domain(s)} & \# \textbf{Kernels} \\
\midrule
HeCBench~\cite{Jin2023HeCBench} & open-source benchmark suite spanning 10 application domains (ML/DNN,
etc.) & All 10 Domains & 770 \\
cuBLAS         & closed-source BLAS library kernels from NVIDIA's cuBLAS SDK & Linear Algebra & 4{,}137 \\
CUTLASS~\cite{cutlass}        & NVIDIA's open-source CUDA Templates for Linear Algebra Subroutines
& Linear Algebra, ML/DNN, Science, Infra., Other & 641 \\
FlashAttention\protect\footnotemark~\cite{flashattention} & optimized attention kernels from the FlashAttention library
& ML/DNN & 63 \\
cuDNN CNN~\cite{cudnn}      & closed-source deep learning library. Convolutional-neural-network kernels & ML/DNN & 46 \\cuDNN Ops      & general tensor-operation kernels from cuDNN's Ops sub-library & ML/DNN & 2{,}283 \\
cuDNN Adv      & machine-learning kernels from cuDNN's Adv sub-library & ML/DNN & 4{,}018 \\
CUDA SDK~\cite{cudasdk}       & CUDA kernels from CUDA SDK examples extracted via cuObjDump~\cite{cuda_binary_utilities} & ML/DNN, Sort/Search, Infra., Other & 19 \\
\midrule
\textbf{Total} & & & \textbf{11{,}977} \\
\end{tabularx}
\end{table*}
\footnotetext{The 63 FlashAttention kernels form a
curated set spanning forward and backward passes, head dimensions
64--256, and FP16/BF16/FP8 precisions.}

\begin{table}[t]
\caption{Distribution of 11{,}977 benchmark kernels}
\label{tab:domain_distribution}
\footnotesize
\begin{center}
\begin{tabular}{l|r}
\textbf{Application Domain} & \textbf{Count} \\%\hline\hline
\midrule
ML/DNN           & 6{,}552   \\
Linear Algebra   & 4{,}499   \\
Infrastructure   & 251     \\
Other            & 250     \\
Science          & 141     \\
Sort/Search      & 109     \\
Simulation       & 80      \\
Image/Vision     & 50      \\
Stencil/PDE      & 21      \\
Graph            & 17      \\
Crypto           & 7       \\%\hline
\midrule
\textbf{Total}   & \textbf{11{,}977} \\
\end{tabular}
\end{center}
\vspace{-4mm}
\end{table}

%%CHANGE: Renamed from "Execution Results" to "Lift Results"
\subsection{Results and Limitations}
\label{sec:eval:execution}

\name{} successfully lifts all kernels across all eight suites to valid LLVM IR (i.e., they pass the LLVM verifier's checks). Correctness is then verified against a native reference: we use bit-exact comparison for the vendor libraries (cuBLAS, cuDNN) and relative-tolerance comparison for the remaining suites. Each kernel binary lifted to LLVM IR is retargeted to the CPU backend and checked against the native reference outputs. We verify end-to-end correctness for all 229 HeCBench applications and 105 CUTLASS examples. For the closed-source libraries, we inject synthesized inputs and execute each kernel on a GPU to obtain an oracle. A kernel is non-observable if it exposes no pointer parameter for an output buffer, or if the original binary faults on the synthesized inputs. Among all lifted kernels, the observable set narrows to 3{,}584 kernels with an executable oracle. Execution verification yields 3{,}452 passing kernels (96.3\%). Key causes of the failed kernels include approximation errors in transcendental functions and floating-point reduction operations.

% Also, the NVVM backend needs to fully map various instruction patterns and new instruction variants to the appropriate NVVM internal functions. We found that, due to the lack of complete documentation, this requires a lot of trial and error.

% We believe the gap in execution correctness stems from the backend implementation that requires more engineering effort, rather than an algorithmic change. While our lifters produce corresponding IR-level output from the SASS input by handling all possible SASS-level instructions, the exact behavior of instructions unique to the GPU architecture is determined by our backend. A correct execution model in the CPU backend requires coordinating individual CPU threads to collectively emulate warp and warp-group-like behavior. We have proven that careful implementation can achieve exact behavior, as features such as warp operations and earlier generations of tensor cores are already implemented(see section~\ref{sec:usecases:x86}). The missing implementation gap remains in deterministically reproducing the behavior for the Hopper features, including the latest generation of tensor core instructions and TMA. In addition, the NVVM backend needs a comprehensive mapping from various instruction patterns and variants of new instructions to the appropriate NVVM intrinsics. We found it required extensive trial and error due to the lack of complete documentation.

\subsection{Type Propagation Characterization}
\label{sec:eval:typechar}

We validate the type-recovery claims of
Section~\ref{sec:hard:types} by measuring six metrics in a
single analysis pass over all 11{,}977 kernels, so every figure
in this section describes the same population.

% \textbf{Instruction classification.}
% Figure~\ref{fig:type_classification} shows the fraction of
% instructions in each recovery role per domain. Type-seeding
% instructions dominate across all suites (95.6\% combined,
% %%CHANGE: Removed per-suite number dump; single range instead
% ranging from 77.2\% in the CUDA SDK to 97.1\% in cuDNN Adv).
% Most values receive their type directly from a seed instruction
% without requiring propagation. The CUDA SDK shows a higher
% transparent fraction 22.8\%, reflecting simpler hand-written
% kernels. Although seeding covers most instructions, the
% type-transparent gap is sufficient to break correctness without
% propagation, as the ablation confirms
% (Section~\ref{sec:eval:ablation}).

\begin{figure}[t]
    \centering
    \includegraphics[width=0.9\linewidth]{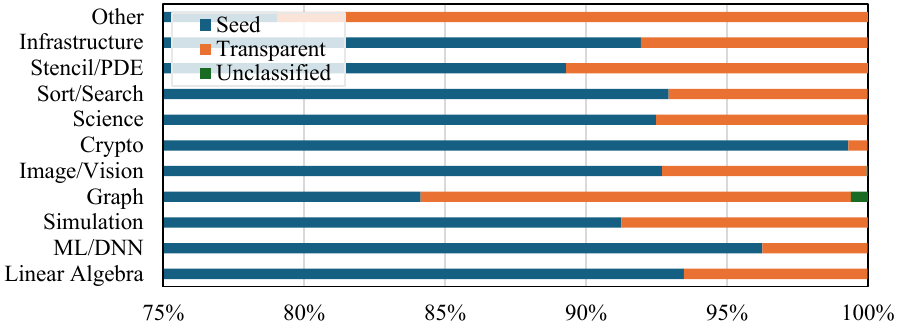}
    \caption{Instruction classification by recovery role per application domain (stacked percentage). All domains show $>$79\% seeding.}
    \label{fig:type_classification}
    \vspace{-3mm}
\end{figure}

\textbf{Instruction classification.}
Figure~\ref{fig:type_classification} shows the fraction of
instructions in each recovery role per domain. Type-seeding
instructions dominate every domain (95.6\% combined, ranging
from 79.1\% in \emph{Other} to 99.3\% in Crypto).
Most values receive their type directly from a seed instruction
without requiring propagation. The domains with the smallest
seeded share also carry the largest type-transparent fraction
(\emph{Other} at 20.9\%, Graph at 15.3\%). Although seeding
covers most instructions, the
type-transparent gap is sufficient to break correctness without
propagation, as the ablation confirms
(Section~\ref{sec:eval:ablation}).

\begin{figure}[t]
    \centering
    \includegraphics[width=0.9\linewidth]{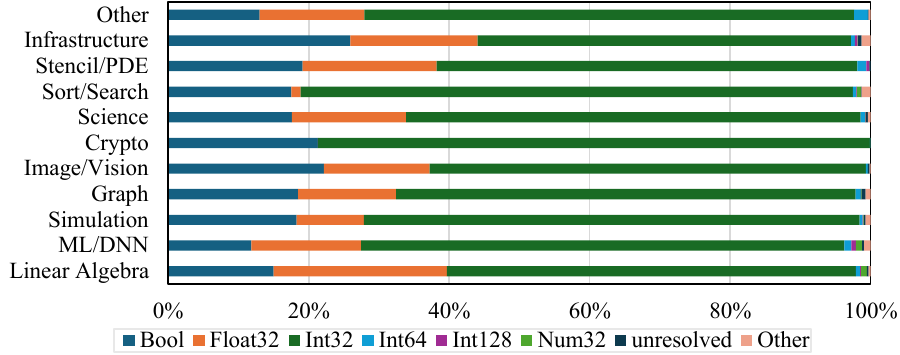}
    \caption{Proportion of type by type recovery across domains.}
    \label{fig:type_composition}
    \vspace{-3mm}
\end{figure}

\begin{figure}[t]
    \centering
    \includegraphics[width=0.9\linewidth]{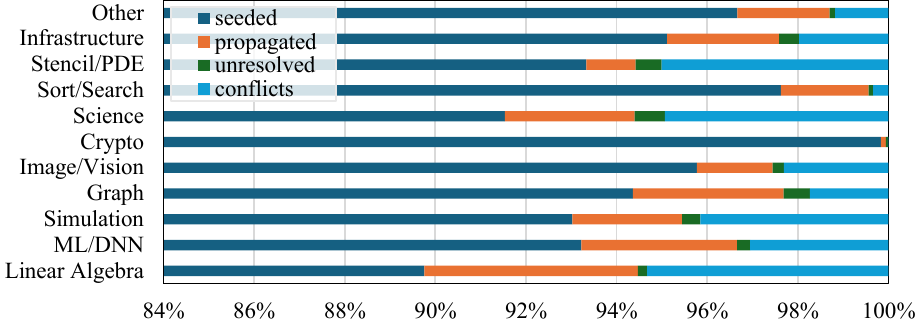}
    \caption{Value resolution by domain (stacked percentage).}
    \label{fig:type_resolution}
    \vspace{-5mm}
\end{figure}

\textbf{Type composition.}
Figure~\ref{fig:type_composition} shows that Int32 is the
dominant type across all domains (67.1\% overall), driven by
addressing and fixed-point computation. Float32 contributes a
substantial share in compute-intensive domains (linear
algebra, ML, science) but is much smaller in integer-heavy
domains such as sort/search and crypto. Int64 accounts for
only a small fraction (0.9\%) because global-memory addressing
operations that require 64-bit pointers are far less frequent
than arithmetic and control instructions. Float16 (0.005\% overall)
appears only in ML/DNN tensor-core kernels. Float64 is
similarly rare, and both fold into \emph{Other} in the figure.
%%CHANGE: Int128 explanation updated to match the full-corpus data
Int128 values (0.56\%, mostly in vendor-library kernels) come
from 128-bit vector memory accesses. SASS has no 128-bit ALU.
A \texttt{LDG.128} or \texttt{STG.128} instruction moves four
32-bit values through one four-register group, and the analysis
types that group as Int128. The lifter lowers each such access
to four 32-bit operations, so Int128 only describes the width
of a memory transaction and never appears as an arithmetic
type in the emitted IR.

\textbf{Type resolution.}
Figure~\ref{fig:type_resolution} breaks down value-resolution
outcomes. Across all suites, \revise{95.9}\% of values are resolved
directly by seed instructions, while propagation resolves
another \revise{3.8}\%, filling the gap left by type-transparent
instructions. Type conflicts constitute \revise{3.7}\% of total values,
and \revise{0.3}\% remain unresolved after propagation, typically
because no type evidence reaches them through the def-use
chain. These unresolved values receive the Int32 fallback
(Section~\ref{sec:typepropagation}).

\begin{figure}[t]
    \centering
    % --- First Subfigure ---
    \begin{subfigure}[b]{0.48\linewidth}
        \centering
        \includegraphics[width=\linewidth]{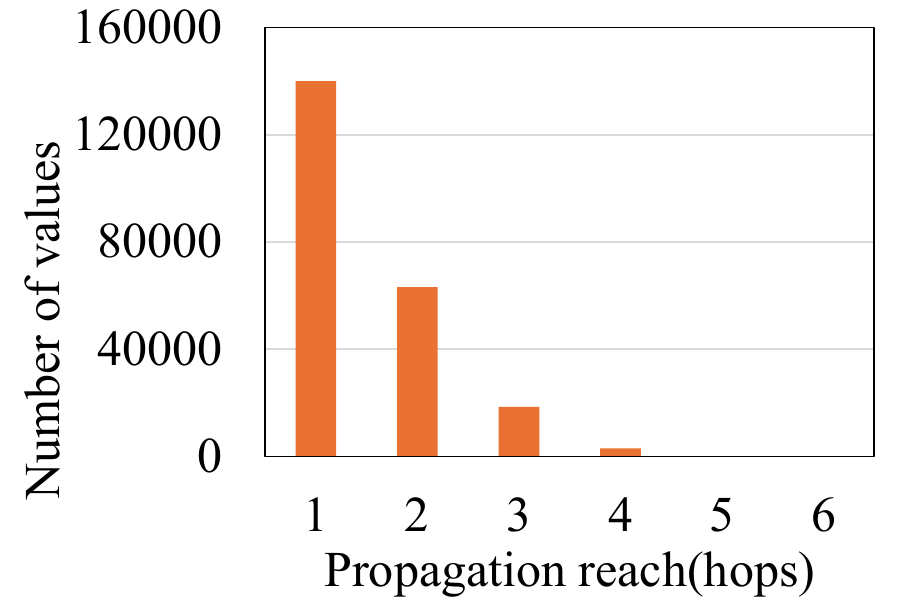}
        \caption{Propagation reach.}
        \label{fig:propagation_reach}
    \end{subfigure}
    \hfill
    % --- Second Subfigure ---
    \begin{subfigure}[b]{0.48\linewidth}
        \centering
        \includegraphics[width=\linewidth]{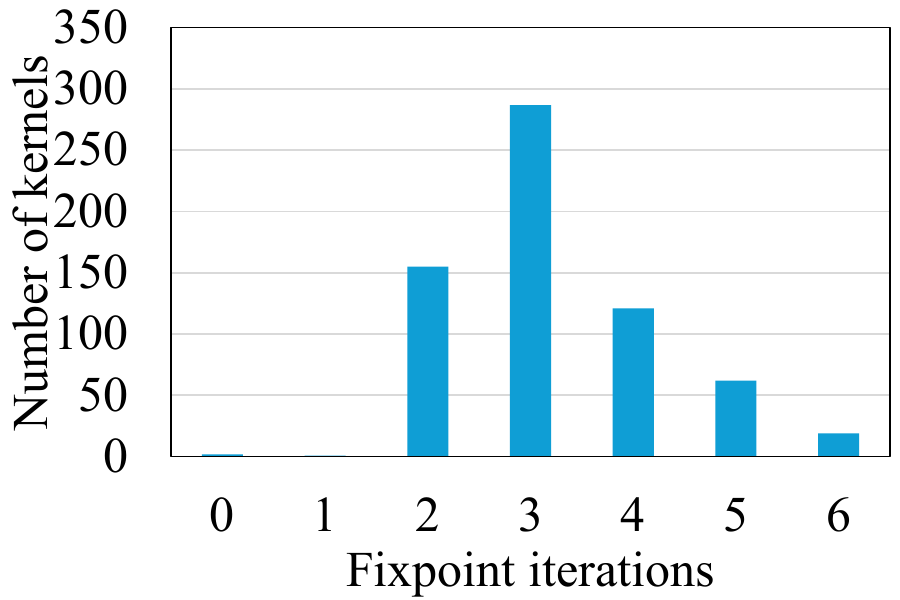}
        \caption{Fixpoint iteration counts.}
        \label{fig:convergence}
    \end{subfigure}
    \caption{Analysis of benchmark propagation chains and analysis convergence (SM90). \textbf{(a)} Distribution of propagation reach (hop count from nearest type-seeding instruction to resolved value) across all benchmarks. Hop=0 (seeded locally) excluded. \textbf{(b)} Distribution of fixpoint iteration counts per lift unit across all benchmarks.}    \label{fig:combined_analysis}
    \vspace{-3mm}
\end{figure}

\textbf{Propagation behavior.}
Figure~\ref{fig:propagation_reach} shows that type evidence
usually travels only 1--2 hops from the nearest seed
instruction, with a small tail extending to 6 hops. Local
evidence is typically sufficient, while a smaller set of cases
requires multi-hop propagation through longer transparent
chains.

%%CHANGE: Added explanation of what iterative analysis means
\textbf{Convergence.}
Each fixpoint iteration propagates type constraints along
def-use edges and checks for new conflicts. The analysis terminates when no type set changes.
Figure~\ref{fig:convergence} shows that convergence is fast in
practice: the mean iteration count ranges from \revise{2.2} (CUDA SDK)
to \revise{4.6} (cuDNN Adv), with a maximum of 6 and a combined mean of
\revise{3.2}. Long propagation chains are uncommon, and the analysis
converges reliably across all benchmark suites.

\subsection{Ablation Study}
\label{sec:eval:ablation}

%%CHANGE: Moved HeCBench scope to top so reader knows immediately
We isolate the contribution of each recovery step by
selectively disabling it and measuring the effect on lifting
success. We focus the primary ablation on HeCBench and confirm generalization on a vendor-library subset.
% We restrict this experiment to HeCBench (229 kernels)
% due to the cost of lifting and executing larger suites.

\begin{figure}[t]
  \centering
  % Top: export WITHOUT x-axis tick labels, x-axis title, or legend
  \begin{subfigure}{\linewidth}
    \centering
    \includegraphics[width=0.9\linewidth]{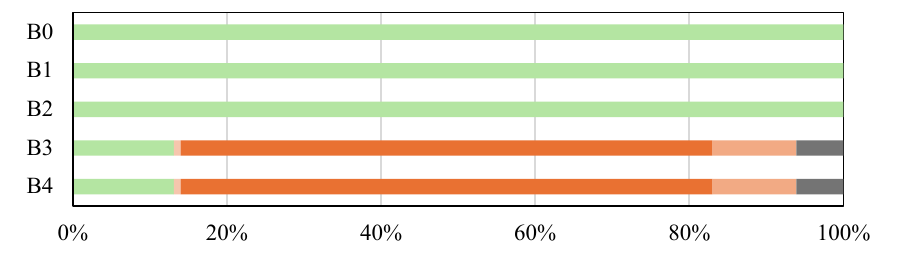}
    \caption{HeCBench}
    \label{fig:ablation-hecbench}
  \end{subfigure}

  \vspace{2pt}

  % Bottom: export WITH the shared x-axis label and the legend baked in
  \begin{subfigure}{0.9\linewidth}
    \centering
    \includegraphics[width=\linewidth]{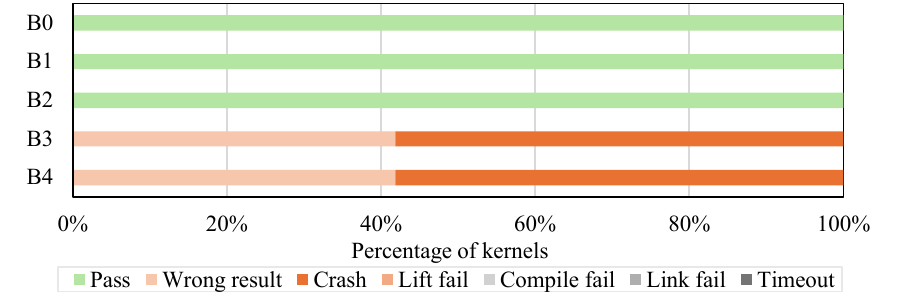}
    \caption{NVIDIA libraries(NPP, NvJPEG)}
    \label{fig:ablation-nvlib}
  \end{subfigure}

  \caption{Ablation on HeCBench and Nvidia Libraries}
  \label{fig:ablation}
\end{figure}

%%CHANGE: Tightened x86 backend explanation -- removed per-thread
%%SIMT state details (already in Section 7.2)
Execution correctness is evaluated via the x86 backend, which
emulates GPU execution for validation
(Section~\ref{sec:usecases:x86}). The full pipeline (B0)
achieves a 100\% x86 pass rate on HeCBench. All ablations run
on the same backend, so any drop from the baseline isolates the
effect of the disabled step.
Table~\ref{tab:ablation_table} and Figure~\ref{fig:ablation-hecbench}
show the results.

% \begin{table}[t]
% \caption{Type recovery steps ablation}
% \label{tab:ablation_table}
% \footnotesize
% \begin{center}
% \begin{tabular}{l|l|r|r|l}
%  & \textbf{What is disabled} & \textbf{IR Gen} & \textbf{x86 Pass} & \textbf{Effect} \\\hline\hline
% B0 & Nothing (full pipeline) & 229 & 169 (73.8\%) & baseline \\
% B1 & Predicate$\to$SEL        & 228 & 167 (72.9\%) & $-$0.9\% \\
% B2 & Pattern normalization     & 229 & 169 (73.8\%) & no change \\
% B3 & Seed only (no propagation)   & 8 & 0 (0\%) & 221 lift fail \\
% B4  & No type recovery (all int32) & 8 & 0 (0\%) & 221 lift fail \\
% \end{tabular}
% \end{center}
% \end{table}

% \begin{table}[t]
% \caption{Type recovery steps ablation}
% \label{tab:ablation_table}
% \footnotesize
% \begin{center}
% \begin{tabular}{l|l|r|r|l}
%  & \textbf{What is disabled} & \textbf{IR Gen} & \textbf{x86 Pass} & \textbf{Effect} \\\hline\hline
% B0 & Nothing (full pipeline) & 229 & 229 (100.0\%) & baseline \\
% B1 & Predicate$\to$SEL        & 229 & 228 (99.6\%) & $-$0.4\% \\
% B2 & Pattern normalization     & 229 & 229 (100.0\%) & no change \\
% B3 & Seed only (no propagation)   & 204 & 34 (14.8\%) & $-$85.2\% \\
% B4  & No type recovery (all int32) & 204 & 33 (14.4\%) & $-$85.6\% \\
% \end{tabular}
% \end{center}
% \end{table}

\begin{table}[t]
\caption{Type recovery steps ablation on HeCBench and the NVIDIA libraries}
\label{tab:ablation_table}
\scriptsize
\centering
\begin{tabular}{l|l|r|r|l}
 & \textbf{What is disabled} & \textbf{IR Gen} & \textbf{x86 Pass} & \textbf{Effect} \\ %\hline\hline
 \midrule
\multicolumn{5}{l}{\textbf{HeCBench (229)}} \\ \hline
B0 & Nothing (full pipeline)      & 229 & 229 (100.0\%) & baseline \\
B1 & Predicate$\to$SEL            & 229 & 229 (100.0\%) & no change \\
B2 & Pattern normalization        & 229 & 229 (100.0\%) & no change \\
B3 & Seed only (no propagation)   & 204 & 30 (13.1\%)   & $-$86.9\% \\
B4 & No type recovery (all int32) & 204 & 30 (13.1\%)   & $-$86.9\% \\ \hline
\multicolumn{5}{l}{\textbf{NVIDIA libraries (98)}} \\ \hline
B0 & Nothing (full pipeline)      & 98 & 98 (100.0\%) & baseline \\
B1 & Predicate$\to$SEL            & 98 & 98 (100.0\%) & no change \\
B2 & Pattern normalization        & 98 & 98 (100.0\%) & no change \\
B3 & Seed only (no propagation)   & 98 & 0 (0.0\%)    & $-$100.0\% \\
B4 & No type recovery (all int32) & 98 & 0 (0.0\%)    & $-$100.0\% \\
\end{tabular}
\end{table}

% Huanzhi: horizontally stacked double column version
% \begin{table*}[t]
% \caption{Type recovery steps ablation on HeCBench and Nvidia Libraries}
% \label{tab:ablation_table}
% \footnotesize
% \centering
% \begin{tabular}{l|l|r|r|l|r|r|l}
%  & & \multicolumn{3}{c|}{\textbf{HeCBench (229)}} & \multicolumn{3}{c}{\textbf{NVIDIA libraries (98)}} \\
%  \cline{3-8}
%  & \textbf{What is disabled} & \textbf{IR Gen} & \textbf{x86 Pass} & \textbf{Effect}
%                              & \textbf{IR Gen} & \textbf{x86 Pass} & \textbf{Effect} \\ \hline\hline
% B0 & Nothing (full pipeline)      & 229 & 229 (100.0\%) & baseline  & 98 & 98 (100.0\%) & baseline \\
% B1 & Predicate$\to$SEL            & 229 & 228 (99.6\%)  & $-$0.4\%  & 98 & 98 (100.0\%) & no change \\
% B2 & Pattern normalization        & 229 & 229 (100.0\%) & no change & 98 & 98 (100.0\%) & no change \\
% B3 & Seed only (no propagation)   & 204 & 34 (14.8\%)   & $-$85.2\% & 98 & 0 (0.0\%)    & $-$100.0\% \\
% B4 & No type recovery (all int32) & 204 & 33 (14.4\%)   & $-$85.6\% & 98 & 0 (0.0\%)    & $-$100.0\% \\
% \end{tabular}
% \end{table*}

% \begin{figure}[t]
%     \centering
%     \includegraphics[width=1.0\linewidth]{figs/ablation_4.pdf}
%     \caption{Ablation on HeCBench (229 kernels, x86 backend). Removing type recovery (B3, B4) causes \revise{85\%} of kernels to fail IR generation. \revise{Other ablations have minimal impact to the baseline}. NVVM pass rate is 99.6\% (Table~\ref{tab:passrates}).}
%     \label{fig:ablation}
% \end{figure}

\textbf{Type propagation gates lifting.}
When propagation is disabled (B3) or when type recovery is
removed entirely (B4), the x86 pass rate collapses from 100\%
to $13.1\%$. Lifting itself still succeeds on 204 of 229
benchmarks, but forcing every value to Int32 reinterprets the
floating-point operations as integer arithmetic. The resulting
binaries compute wrong addresses and values and crash at
runtime (158 of 229 segfault), with the remaining 25 benchmarks
failing to lift where the missing type information leaves
operand widths unresolvable. Disabling predicate-to-select
lowering (B1) or pattern normalization (B2) has negligible
effect on lift success, confirming that type propagation is the
single step that gates correct lifting.

\begin{figure}[t]
    \centering
    \includegraphics[width=0.85\linewidth]{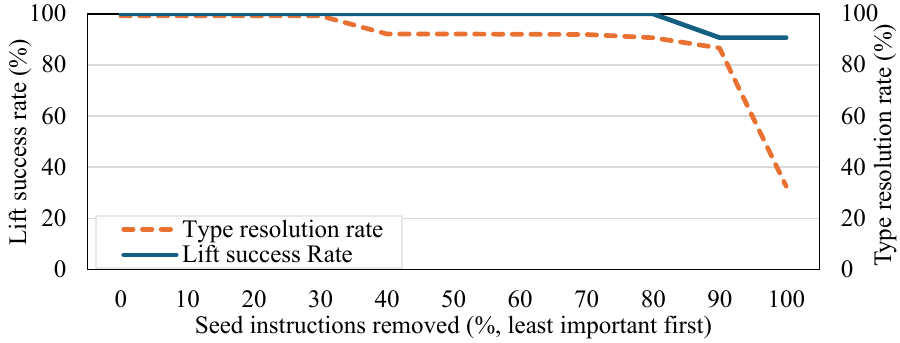}
    \caption{Type degradation on HeCBench. Lift success and type resolution remain stable under modest seed removal, then drop sharply once too much seed is removed.}
    \label{fig:type_degradation}
\end{figure}

\textbf{Resilience to reduced seed coverage.}
Figure~\ref{fig:type_degradation} progressively removes seed
instructions in ascending order of frequency. Lift success and
type resolution remain stable under modest seed loss, then
degrade sharply once too many seeds are removed, showing that
propagation compensates for moderate seed loss but depends on
a core set of high-frequency seed instructions.

\textbf{Generalization across workload classes.}
We repeat the ablation on 98 kernels from two NVIDIA libraries outside our main suites, NPP (image and signal processing) and nvJPEG (image codecs).
% These workloads are dominated by integer and bit-manipulation code rather than the floating-point linear algebra of cuBLAS, cuDNN, and CUTLASS, giving the ablation a substantially different instruction and type mix to test against. 
Figure~\ref{fig:ablation-nvlib} and Table~\ref{tab:ablation_table} report the result. The full pipeline executes every kernel correctly, B1 and B2 leave lifting unchanged, and disabling propagation (B3) or type recovery (B4) causes every kernel to execute incorrectly. 
% Type recovery gates lifting on these libraries exactly as on
% HeCBench, confirming the conclusion is not an artifact of a single benchmark
% suite or workload class.

\subsection{Type Boundary Analysis}
\label{sec:eval:typeboundary}

%%CHANGE: Added opening definition of type boundaries
Type boundaries are program points where constraint propagation
detects an empty intersection, indicating that the compiler has
reinterpreted a register's bits as a different type without an
explicit conversion. By analyzing the instructions at these
boundaries, we find that most are not due to incorrect type
recovery but rather intentional compiler optimizations.
Table~\ref{tab:conflict_pattern_table} summarizes the main
categories: bit-level manipulation of floating-point values
(IEEE~754 field extraction, float reconstruction), fast-math
approximation chains (integer reciprocal via
\texttt{I2F}$\to$\texttt{MUFU.RCP}$\to$\texttt{IADD3}$\to$%
\texttt{F2I}), and branch merges that join semantically
distinct type paths.

\begin{figure}[t]
    \centering
    \includegraphics[width=0.9\linewidth]{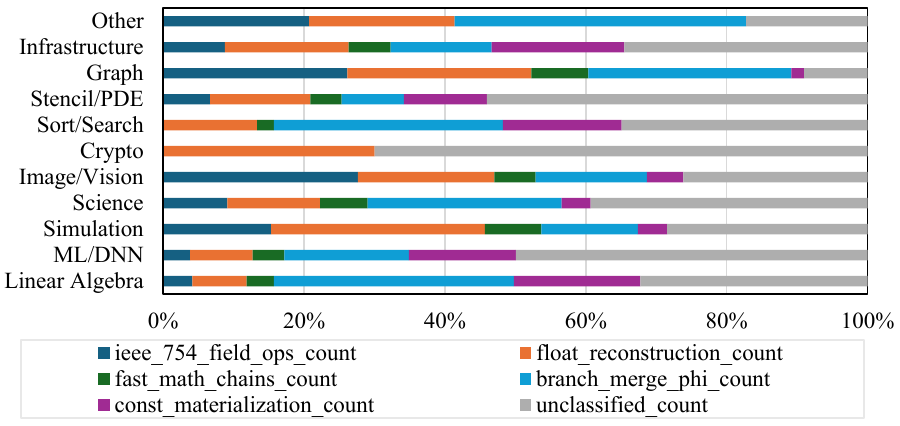}
    \caption{Distribution of instruction patterns at type boundaries per application domain (stacked percentage).}
    \label{fig:conflict_taxonomy}
\end{figure}

Figure~\ref{fig:conflict_taxonomy} categorizes the boundary
patterns across all domains.
%%CHANGE: "significant share" → "majority" (Rule 7: be specific)
% Classifiable patterns account for the majority of boundaries in
% all domains except Crypto, which contains no type boundaries
% because its kernels use almost exclusively integer arithmetic.

\begin{table}[t]
\caption{Type boundary instruction patterns categorization.}
\label{tab:conflict_pattern_table}
\footnotesize
\centering
{
\setlength{\tabcolsep}{5pt}
\renewcommand{\arraystretch}{1.08}
\begin{tabular}{p{2cm} | p{5cm}}
%\toprule
\textbf{Category} & \textbf{Example pattern} \\
\midrule
Branch merge
& a \texttt{PHI} node joins a float path with an integer path (LOP3 bit-manipulation). \\
\midrule
Fast math chains
& \texttt{I2F} $\rightarrow$ \texttt{MUFU.RCP} $\rightarrow$ \texttt{IADD3} $\rightarrow$ \texttt{F2I}.  Fast integer division. \\
\midrule
IEEE 754 field ops
& \texttt{mantissa\_extraction}: \texttt{LOP3.LUT Rd, Rsrc, 0x7fffff, RZ}. Isolate mantissa field. \\
\midrule
Float reconstruction
& \texttt{LOP3.LUT R53, R43, 0x7f800000, RZ,
0xfc}. Forces Inf/NaN via exponent saturation \\
\midrule
Analysis artifacts & \texttt{IMAD.WIDE} produces a 64-bit result, but downstream only uses the 32-bit low as float. \\
\midrule
Unclassified
& Uncategorized type conflicts. \\
%\bottomrule
\end{tabular}
}
\end{table}

\subsection{Cost of Lifting}
\label{sec:eval:cost}

% Data: scripts/measure_lift_cost.py --suite all -> eval_stats/lift_cost_full/
% Figure + CSV: scripts/gen_cost_figure.py -> figs/lift_cost.pdf, figs/csv/figC_lift_cost.csv
We measure cost on the same eight-suite corpus used for
correctness (Table~\ref{tab:passrates}). The lift unit is the
cubin, where lifting one cubin recovers every kernel it
contains. The 650 cubins yield 11{,}977 lifted kernel definitions
plus their embedded device helpers. Each cubin is lifted in a
fresh process on a single CPU core (pure-Python, CPython~3.12,
\texttt{llvmlite}~0.48/LLVM~20). We record wall-clock time for
the full pipeline and peak resident memory.

\begin{table}[t]
\caption{Lifting cost per benchmark suite. \emph{med.}/\emph{max}~$t$ are per-cubin lift
times. \emph{RSS} is peak resident memory.}\label{tab:lift_cost}
\scriptsize
\centering
\renewcommand{\arraystretch}{1.1}
\setlength{\tabcolsep}{3pt}
\begin{tabular}{l|r|r|r|r|r|r}
\textbf{Suite} & \textbf{kernels} & \textbf{cubins} & \textbf{insts} (M) & \textbf{med.\,$t$} (s) & \textbf{max\,$t$} (s) & \textbf{RSS} (GiB) \\
 % \hline
 \midrule
HeCBench       & 770 & 229 & 0.37 &   0.50 &  53.9 & 0.7 \\
cuBLAS         & 4{,}137 & 177 & 5.59 &   3.69 & 565.8 & 12.0 \\
CUTLASS        & 641 & 105 & 0.72 &   2.31 &  96.0 & 2.1 \\
FlashAttention &  63 &  63 & 7.03 &  55.93 & 148.8 & 3.0 \\
cuDNN CNN      &  46 &   7 & 0.01 &   0.58 &   3.4 & 0.1 \\
cuDNN Ops      & 2{,}283 &  20 & 2.99 &  28.97 & 422.6 & 6.3 \\
cuDNN Adv      & 4{,}018 &  30 & 6.49 & 164.65 & 470.6 & 9.3 \\
CUDA SDK       &  19 &  19 & 0.00 &   0.03 &   0.3 & 0.1 \\
% \hline
% \textbf{Total} & \textbf{1{,}681} & \textbf{650} & \textbf{23.2} & \multicolumn{3}{c}{} \\
\end{tabular}
\end{table}

% \begin{figure*}[t]
% \centering
% \includegraphics[width=0.9\linewidth]{figs/lift_cost_2.pdf}
% \caption{Lifting cost vs.\ kernel size, one point per cubin. \textbf{(a)}~Total lift time scales near-linearly
% with instruction count. \textbf{(b)}~Peak memory scales linearly
% ($\approx\!17.6$\,KiB per instruction); 
% % the $\sim$50\,MiB floor is the Python
% % interpreter baseline. Vendor-library kernels (cuBLAS, cuDNN), absent from
% % HeCBench, populate the high end ($>10^5$ instructions) and confirm the trend
% % holds across 4.5 orders of magnitude.
% }
% \label{fig:lift_cost}
% \end{figure*}

\noindent\textbf{Scaling.}
Lifting cost is linear in code size over the full range.
Table~\ref{tab:lift_cost} show
that the entire corpus lifts in under five hours on a single
core, at a median of $1.8$\,s per cubin. A least-squares fit
gives $0.83$\,ms of lift time and $17.6$\,KiB of peak memory per
SASS instruction. The worst
single cubin from our benchmark set is a $507{,}656$-instruction cuDNN~Adv binary. 
% It takes $9.4$\,min and $9.5$\,GiB. The largest memory footprint is $12$\,GiB on a cuBLAS cubin, which stays well within commodity RAM.
% Within a given size, another cost driver is control flow. Lift time correlates more strongly with basic-block count ($r{=}0.84$) than with raw instruction count ($r{=}0.55$),
% because heavily predicated code is split into many small blocks
% and the per-block passes (SSA renaming, dead-code elimination,
% IR emission) dominate. Parsing is under $10\%$ of
% lift time, while transforms and IR emission account for the rest.

\noindent\textbf{Type-inference cost.}
Each iteration of type analysis (Section~\ref{sec:eval:typechar}) is a single $O(\text{instructions})$ sweep, and the number of iterations stays a small bounded constant. The
per-function mean rises only from $2.3$ on the smallest kernels
to $4.2$ on the largest. A $1000\times$ increase in size
yields under a $2\times$ increase in iterations. It is dominated by the
control-flow passes.

\section{Use Cases}
\label{sec:usecases}

\subsection{Memory Access Pattern Analysis}

We demonstrate analysis of closed-source kernels on a kernel with stride-based thread-variant memory accesses, a common pattern in ML and HPC code that causes poor coalescing. The SASS below, recovered from a compiled binary, shows the first violation: non-contiguous accesses where Thread~0 reads \texttt{input[0]}, Thread~1 reads \texttt{input[5]}, Thread~2 reads \texttt{input[10]}, and so on, forcing 32 memory transactions per warp instead of 1--2:

\begin{keeplisting}
\begin{lstlisting}[]
CGPart_Rank         R3, R0
00b0: ISETP.GE.AND  P1, PT, R6, c[0x0][0x170], PT
00c0: IADD3         R3, R0, R3, RZ
00f0: MOV           R4, 0x4
0100: IMAD.WIDE.U32 R2, R3, R4, c[0x0][0x160]
0110: LDG.E.SYS     R2, [R2]
\end{lstlisting}
\end{keeplisting}

The second case accesses every other element (Thread~0: \texttt{input[0]}, Thread~1: \texttt{input[4]}, ...), requiring 16 transactions for 16 threads:

\begin{keeplisting}
\begin{lstlisting}[]
CGBlock_Rank    R2, R6
0180: MOV       R5, 0x4
0190: IMAD.WIDE R2, R6, R5, c[0x0][0x160]
01a0: LDG.E.SYS R2, [R2]
01b0: IMAD.WIDE R4, R6, R5, c[0x0][0x168]
01c0: FADD      R7, R2, 1
01d0: STG.E.SYS [R4], R7
\end{lstlisting}
\end{keeplisting}

Both cases can be identified from lifted IR by tracing
def-use chains from thread indices through address
computations. This analysis requires typed IR: without type
recovery, the lifter cannot emit valid LLVM IR at all
(Section~\ref{sec:eval:ablation}), and even partially
incorrect types would misclassify integer index arithmetic as
floating-point operations, breaking the address pattern.

% Huanzhi: the below example does not show why BSYNC is redundant.
% We also identify unnecessary synchronization. The following SM90 SASS retains a \texttt{BSYNC} from \texttt{tile.sync()} that our lifted-IR analysis shows to be redundant:

% \begin{lstlisting}[]
% 01f0: @!P1 FADD     R5, R5, R8
% 0200: SHFL.DOWN     PT, R4, R5, 0x1, 0x1c1f
% 0210: LOP3.LUT      P1, RZ, R3, 0x3, RZ, 0xc0, !PT
% 0220: @!P1 FADD     R5, R5, R4
% 0230: BSYNC         B0
% \end{lstlisting}

\subsection{Cross-Platform Execution on x86 and ARM}
\label{sec:usecases:x86}

Because the lifted output is standard LLVM~IR, we can retarget
it to non-GPU backends including x86-64 and aarch64 (ARM). We
compile lifted IR with \texttt{clang} and execute on
commodity hardware, enabling functional testing without a GPU.

\textbf{Execution model.} The CPU runtime emulates GPU execution for correctness
validation, not performance: one OS~thread per GPU logical
thread, with CTAs executed sequentially. Kernel parameters are
passed through constant memory matching SASS's
\texttt{c[0][ARG\_BASE + offset]} layout, so the lifted
function has a \texttt{void()} signature with no ABI
dependency. The base offset varies across GPU generations
(e.g., \texttt{0x160} on SM75, \texttt{0x210} on SM90,
\texttt{0x360} on SM100/SM120) because NVIDIA expands the
constant-memory prefix with each generation to accommodate
additional system state. \name{} reads the correct offset from
ELF metadata (\texttt{EIATTR\_PARAM\_CBANK}) rather than
hardcoding it, so the same runtime supports all architectures. Table~\ref{tab:exec-model} summarizes the full mapping.
% GPU-specific operations have no direct CPU equivalent;
% Table~\ref{tab:gpu-decomp} summarizes how we decompose them to
% portable scalar operations. All decompositions use standard
% LLVM instructions and C++17 synchronization primitives, with
% no platform-specific intrinsics or inline assembly.

%Because the lifted output is standard LLVM~IR, we can retarget it to non-GPU backends including x86-64 and aarch64 (ARM). The entire lifting pipeline (parse, transform, type analysis, IR generation) is target-independent; only the final compilation step selects the CPU target. We compile lifted IR with \texttt{clang~-O2} and execute on commodity x86 or ARM hardware, enabling functional testing without a GPU.

%\textbf{Execution model.}
%The CPU runtime emulates GPU execution for correctness
%validation, not performance, one OS~thread per GPU logical
%thread, with per-thread SIMT state (TID, CTAID, LANE\_ID) stored in thread-local constant memory matching SASS's byte-aligned layout. Kernel parameters are passed through constant memory (matching SASS's \texttt{c[0][ARG\_BASE + offset]} layout) rather than CPU calling conventions, so the lifted function has a \texttt{void()} signature with no ABI dependency. CTAs execute sequentially, and shared memory is backed by a per-CTA global array zeroed between iterations. Table~\ref{tab:exec-model} summarizes the full mapping.

\begin{table}[htbp]
\footnotesize
\centering
\caption{GPU-to-CPU execution model mapping.}
\label{tab:exec-model}
\footnotesize
\begin{tabular}{@{}l|l@{}}
%\toprule
\textbf{GPU Concept} & \textbf{CPU Mapping} \\
\midrule
Grid (gridDim CTAs)       & Sequential CTA loop \\
CTA (blockDim threads)    & \texttt{std::thread} pool (1 per GPU thread) \\
Warp (32 lanes)           & 32 OS threads + mutex-based shuffle/vote \\
Shared memory (per-CTA)   & Global array (zeroed per CTA) \\
Local memory (per-thread) & Thread-local storage (TLS) array \\
Constant memory (5 banks) & TLS array (copied per thread) \\
Registers (R0--R255)      & SSA values (register-allocated by LLVM) \\
Predicates (P0--P7)       & \texttt{i1} values in LLVM IR \\
%\bottomrule
\end{tabular}
\end{table}

\textbf{GPU semantic decomposition.}
GPU-specific operations have no direct CPU equivalent. We decompose them to portable scalar operations as summarized in
Table~\ref{tab:gpu-decomp}. All decompositions use standard
LLVM instructions and C++17 synchronization primitives, with
no platform-specific intrinsics or inline assembly.

% \begin{table}[htbp]
% \footnotesize
% \centering
% \caption{GPU semantic decomposition to portable CPU operations.}
% \label{tab:gpu-decomp}
% \footnotesize
% \begin{tabular}{@{}lll@{}}
% \toprule
% \textbf{GPU Operation} & \textbf{CPU Decomposition} & \textbf{Notes} \\
% \midrule
% HMMA/WGMMA & Scalar FMA loops  & Correct for lane~0 \\
% (tensor core) &  &  \\
% SHFL (warp shuffle)       & Mutex-based  & Per-warp \\
%   & exchange  & shared buffer \\
% VOTE/BALLOT  & Mutex-based  & Per-warp \\
% (warp vote)   & ballot  & ballot buffer \\
% BAR (barrier)             & Condition-variable wait & Counting barrier \\
% ATOM (atomics)            & Scalar load+op+store & Sequential CTA  \\
% & & model \\
% MUFU (special math)       & \texttt{libm} functions & +3~ULP bias for RCP \\
% MEMBAR/DEPBAR (fences)    & NOP & Ordering via mutex \\
% LDG/STG (global mem)      & \texttt{inttoptr} + load/store & \\
% LDS/STS (shared mem)      & GEP into shared array & \\
% \bottomrule
% \end{tabular}
% \end{table}

\begin{table}[htbp]
\scriptsize
\centering
\caption{GPU semantic decomposition to portable CPU operations.}
\label{tab:gpu-decomp}
\begin{tabular}{@{}l|l|l@{}}
%\toprule
\textbf{GPU Operation} & \textbf{CPU Decomposition} & \textbf{Notes} \\
\midrule
\makecell[l]{HMMA/WGMMA {\scriptsize (tensor)}} % core)}}
  & Scalar FMA loops
  & Correct for lane~0 \\

\makecell[l]{SHFL {\scriptsize (warp shuffle)}}
  & \makecell[l]{Mutex-based exchange}
  & \makecell[l]{Per-warp shared buffer} \\

\makecell[l]{VOTE/BALLOT {\scriptsize (warp vote)}}
  & \makecell[l]{Mutex-based ballot}
  & \makecell[l]{Per-warp ballot buffer} \\

BAR {\scriptsize (barrier)}
  & Condition-variable wait
  & Counting barrier \\

ATOM {\scriptsize (atomics)}
  & Scalar load+op+store
  & \makecell[l]{Sequential CTA model} \\

MUFU {\scriptsize (special math)}
  & \texttt{libm} functions
  & +3~ULP bias for RCP \\

MEMBAR/DEPBAR {\scriptsize (fences)}
  & NOP
  & Ordering via mutex \\

LDG/STG {\scriptsize (global mem)}
  & \texttt{inttoptr} + load/store
  & \\

LDS/STS {\scriptsize (shared mem)}
  & GEP into shared array
  & \\
%\bottomrule
\end{tabular}
\end{table}

\textbf{Numerical fidelity.}
The hardest semantic gap is NVIDIA's hardware reciprocal (\texttt{MUFU.RCP}), which returns an approximation biased by approximately $+3$~ULPs relative to IEEE~754. The compiler-emitted integer-division sequence depends on this bias: our runtime must bias the IEEE reciprocal result by $+3$~ULPs to match, or exact IEEE division produces wrong quotients.

\textbf{ARM (aarch64) support.}
Supporting ARM required changing only the TLS model annotation
and the LLVM target triple. The entire lifting pipeline
(parser, transform passes, type analysis, and IR generation) is
target-independent and produces identical LLVM~IR regardless of
the host CPU. We validated on an ARM Grace Hopper (GH200) system. All failures matched x86 failures with no ARM-specific issue.  ARM's relaxed memory model does not affect correctness because runtime inter-thread communication uses \texttt{std::mutex},
which provides acquire/release semantics on both
architectures.

The goal of cross-platform execution is analysis and validation, not performance parity. Texture/surface operations and SM90 cluster-level features are not yet supported.

%% ─────────────────────────────────────────────────────────────────────────

\section{Related Work}
\label{sec:related}

\subsection{CPU Binary Lifters and Their Assumptions}

CPU binary lifters, including McSema~\cite{mcsema}, BAP~\cite{cpu_binary:2011:bap}, RetDec~\cite{retdec}, and others~\cite{cpu_binary:2000:uqbt, cpu_binary:2012:sp, cpu_binary:2017:ase, cpu_binary:2020:pldi, cpu_binary:2020:sp, cpu_binary:2008:vx32}, are mature tools for x86, ARM, and RISC-V. As discussed in Section~\ref{sec:hard}, the assumptions these
tools rely on (branch-based control flow, separate register
files, documented ISAs) do not hold for SASS.

\subsection{Type Recovery: CPU vs. GPU}
CPU binary \typerecov{} is well-studied, with two main lines of work.
The first is constraint-based inference:
TIE~\cite{tie} introduced type inference with intersection and union types,
Retypd~\cite{retypd} advanced this direction with recursive types, polymorphism, and subtyping,
and BinSub~\cite{binsub2024} reformulates subtyping-based inference using algebraic subtyping, achieving a 63$\times$ runtime improvement over Retypd.
The second line applies probabilistic and learning-based techniques:
OSPREY~\cite{osprey} recovers variables and data structures probabilistically,
DEBIN~\cite{debin} and DIRTY~\cite{dirty2022} predict types and variable names from stripped binaries using learning-based models,
Manta~\cite{manta2024} combines hybrid-sensitive static analysis with type inference for bug detection,
and TYGR~\cite{tygr} infers types for stripped binaries via graph neural networks.
All of these systems operate in settings where ISA and ABI structure provide strong type evidence through separate register files, explicit branch instructions, and documented calling conventions.

On SASS, that evidence is much weaker because integer, floating-point, pointer, and packed values all share a single architectural register file.
GPU \typerecov{} therefore becomes a propagation problem with conflict detection rather than a classification problem.
We must also handle \emph{intentional} type punning, routinely emitted by the GPU compiler for fast reciprocal and exponent manipulation, through explicit bitcast insertion at conflict boundaries.

%CPU binary type recovery is well studied, spanning constraint-based, probabilistic, learning-based, and neural approaches~\cite{tie, retypd, binsub2024, osprey, debin, dirty2022, manta2024, tygr}. These systems assume settings where ISA and ABI structure provide strong type evidence. On SASS, that evidence is much weaker because integer, floating-point, pointer, and packed values all share a single architectural register file. Consequently, GPU \typerecov{} is less a classification problem than a propagation problem with conflict detection, and it must also handle intentional type punning through explicit bitcast insertion.

\subsection{Tools for SASS or GPU Binary Analysis}

NVIDIA's \textit{cuObjDump} and \textit{nvdisasm}~\cite{cuda_binary_utilities} provide disassembly and CFG extraction. NVBit~\cite{nvbit} is a dynamic binary instrumentation framework. SASSI~\cite{sassi} is a compile-time instrumentation framework. Ari et al.~\cite{cgo:ari:2019} demonstrate assembler generation with a Hardware Abstraction Layer. Unlike \name{}, none of these recover analyzable LLVM IR from SASS.

%\subsection{GPU Binary Analysis}

Jeanmougin et al.~\cite{jeansrc23} build CFGs for SASS using abstract interpretation to detect divergence. Shoushtary et al.~\cite{shoushtary2024} reverse-engineer 20 SASS control-flow instructions for CFG construction. Sen et al.~\cite{SenVS23} apply pattern matching to SASS for memory-access analysis. In contrast, \name{} translates SASS to typed LLVM IR, enabling compiler analysis and retargeting to other architectures.

\subsection{Running CUDA on Non-NVIDIA Hardware}

Prior work translates CUDA at various abstraction levels~\cite{mcuda, FCUDA, DBLP:journals/corr/abs-2112-10034, han2024atdaes, SoftCUDA, Volt, ocelot, zluda, FlexGrip, HaunzhiWorkshop}. Source-level tools (MCUDA~\cite{mcuda}, Hipify~\cite{hipify}) convert CUDA source to CPU loops or HIP. COX~\cite{DBLP:journals/corr/abs-2112-10034} and CuPBoP/SoftCUDA/Volt~\cite{han2024atdaes, SoftCUDA, Volt} operate on NVVM IR. Ocelot~\cite{ocelot} lifts PTX to LLVM IR. ZLUDA~\cite{zluda} translates PTX at runtime. FlexGrip~\cite{FlexGrip} reverse-engineers SASS for integer-only FPGA execution. CASS~\cite{arxiv:25:CASS} ports CUDA to AMD at source and assembly level. Unlike all of these, \name{} targets SASS (the final binary) and recovers typed LLVM IR.

%% ─────────────────────────────────────────────────────────────────────────

%% ─────────────────────────────────────────────────────────────────────────
\section{Conclusion}
\label{sec:conc}

We presented \name{}, a framework that recovers analyzable
LLVM IR from NVIDIA SASS binaries. Our key findings are:
\begin{enumerate}
    \item \textbf{Type recovery is the key challenge of GPU
    binary lifting.} SASS's unified register file erases all
    type information. We show that type recovery on GPUs is a
    constraint-propagation problem with conflict detection. Our
    ablation confirms this: disabling type recovery collapses
    the x86 pass rate from 100\% to 13.1\%, while disabling
    other steps has negligible effect.
    \item \textbf{The predicated conversion to SSA adapts known
    techniques.} We apply $\psi$-function-based select insertion    from the IA-64 literature, extended for SIMT-specific
    constructs (predicated \texttt{EXIT}, dual-predicate
    branches, convergence barriers).
    \item \textbf{Practical lifting across GPU generations is
    feasible.} \name{} supports SM75 through SM120 and
    lifts all kernels to valid IR with more than 90\% end-to-end
    execution correctness in every suite, bringing opaque GPU
    binaries back into the compiler-analysis stack.
\end{enumerate}

The lifted LLVM IR enables downstream analysis (memory access
pattern detection) and cross-platform execution on x86 and ARM
CPUs.

%% ─────────────────────────────────────────────────────────────────────────
%\appendix
%\section{Pattern Matching Details}
%\label{sec:appendix:patterns}
%\input{appendix}

%\begin{acks}
%The authors acknowledge the use of large language models for assistance with %coding, writing, figure generation, and table formatting.
%\end{acks}

% MICRO'26 Artifact Appendix -- CuLifter
\appendix

\section{Artifact Appendix}
%%%%%%%%%%%%%%%%%%%%%%%%%%%%%%%%%%%%%%%%%%%%%%%%%%%%%%%%%%%%%%%%%%%%%
\subsection{Abstract}

This artifact provides CuLifter, our tool for lifting NVIDIA SASS binaries
to LLVM IR, together with the benchmark corpus and evaluation scripts used
in the paper. Reviewers can reproduce two main results: the lifting coverage
over the full benchmark corpus (E1), and the data-driven figures of the
paper, Figures 4--10 (E2).

%%%%%%%%%%%%%%%%%%%%%%%%%%%%%%%%%%%%%%%%%%%%%%%%%%%%%%%%%%%%%%%%%%%%%
\subsection{Artifact check-list}

\begin{itemize}

  \item \textbf{Algorithm:}
        Static lifting of NVIDIA SASS to LLVM IR.

  \item \textbf{Artifact contents:}
        CuLifter, benchmark binaries and their SASS, evaluation scripts,
        figure-generation scripts, reference outputs.

  \item \textbf{Benchmarks:}
        HeCBench, cuBLAS, CUTLASS, FlashAttention-3, cuDNN, and CUDA SDK
        samples, totaling 11,977 sm\_90 kernels, plus the NPP/nvJPEG kernels
        used by the Figure~8 ablation.

  \item \textbf{Platform:}
        x86-64 Linux (Ubuntu 22.04.5 LTS).

  \item \textbf{Hardware:}
        8+ CPU cores and 32\,GB RAM recommended (the largest
        vendor-library modules peak at ${\sim}3.5$\,GB per lifting worker).
        Evaluated on a 48-core Intel Xeon E5-2650~v4 @ 2.20\,GHz with
        251\,GB RAM running Ubuntu~22.04.5~LTS.

  \item \textbf{Software:}
        Python 3.12 with \texttt{llvmlite} 0.48, and LLVM/Clang 20.1.8
        with \texttt{g++} 11+ for the execution-based figures. All
        dependencies install from the provided conda environment file.

\item \textbf{Disk space:}
      At least 50\,GB of free disk space is recommended.

\item \textbf{Setup time:}
      Approximately 10 minutes, covering the conda environment and the
      corpus download and extraction.

\item \textbf{Experiment time:}
      Approximately 1 hour for E1 (8 lifting jobs) and 7 hours for E2 on
      the evaluation machine above. E2's cost is dominated by
      characterizing the large cuDNN modules ($\sim$5 of the 7 hours).

  \item \textbf{Output:}
        Console summaries and per-figure CSVs.

  \item \textbf{Availability:}
        Archived on Zenodo. The code is also available on
        \href{https://github.com/gthparch/CuLifter.git}{GitHub}. The
        benchmark archive is downloaded by the setup script.

  \item \textbf{Archival DOI:}
        \href{https://doi.org/10.5281/zenodo.21545827}{10.5281/zenodo.21545827}

  \item \textbf{Data license:}
        Benchmark components remain subject to their original licenses.

\end{itemize}

%%%%%%%%%%%%%%%%%%%%%%%%%%%%%%%%%%%%%%%%%%%%%%%%%%%%%%%%%%%%%%%%%%%%%
\subsection{Description}

\subsubsection{How to access}

The artifact is archived on Zenodo
(\href{https://doi.org/10.5281/zenodo.21545827}{10.5281/zenodo.21545827}).
The code is also available on
\href{https://github.com/gthparch/CuLifter.git}{GitHub}. The setup script
downloads the benchmark data and extracts it at the repository root, and
the README in \texttt{reproduce} describes the evaluation workflow.

\subsubsection{Hardware dependencies}

The experiments run on an x86-64 Linux machine. We recommend at least 8
CPU cores, 32\,GB of RAM, and 50\,GB of free disk space.

\subsubsection{Software dependencies}

Python 3.12 is required, and the provided conda environment installs
it together with \texttt{llvmlite}, LLVM/Clang~20.1.8, and \texttt{llc}.
E1 needs nothing further. E2 additionally uses \texttt{g++} 11 or newer,
which the ablation figures use to compile and run lifted kernels.

\subsubsection{Data sets}

The benchmark archive contains every kernel evaluated in the paper, from
HeCBench, cuBLAS, CUTLASS, FlashAttention-3, cuDNN, and the CUDA SDK,
plus the NPP/nvJPEG ablation inputs. We provide pre-disassembled SASS for
every kernel alongside its original cubin or ELF.

%%%%%%%%%%%%%%%%%%%%%%%%%%%%%%%%%%%%%%%%%%%%%%%%%%%%%%%%%%%%%%%%%%%%%
\subsection{Installation}

{\footnotesize
\begin{verbatim}
git clone \
  https://github.com/gthparch/CuLifter.git
cd CuLifter
conda env create -f environment.yml \
  -p .conda_env
export PATH="$(pwd)/.conda_env/bin:$PATH"
bash reproduce/setup_data.sh
\end{verbatim}
}

The setup script downloads the benchmark archive, verifies its
checksum, and extracts it to \texttt{benchmark/}.

%%%%%%%%%%%%%%%%%%%%%%%%%%%%%%%%%%%%%%%%%%%%%%%%%%%%%%%%%%%%%%%%%%%%%
\subsection{Experiment workflow}

The complete evaluation runs as a single command:

{\footnotesize
\begin{verbatim}
bash reproduce/run_all.sh
\end{verbatim}
}

This runs both experiments in order and prints a final summary. E1 lifts
every kernel in the corpus with the CPU backend and reports per-suite
coverage. E2 regenerates the paper's data-driven figures (4--10) from the
same corpus. One characterization pass produces Figures~4--7 and~10, and
the execution stages produce Figure~8 and Figure~9.

To run the two experiments separately:

{\footnotesize
\begin{verbatim}
bash reproduce/run_cpu_sm90.sh              # E1
bash reproduce/figures/reproduce_figures.sh # E2
\end{verbatim}
}

%%%%%%%%%%%%%%%%%%%%%%%%%%%%%%%%%%%%%%%%%%%%%%%%%%%%%%%%%%%%%%%%%%%%%
\subsection{Evaluation and expected results}

E1 prints a per-suite table of lifted units and emitted LLVM
definitions, and writes a CSV summary and per-kernel logs to the results
directory printed at the end of the run. The console output we obtained
is pasted in \texttt{reproduce/README.md}.

E2 writes one CSV per figure to \texttt{reproduce/figures/output/},
containing the plotted values of the corresponding paper figure. The
values we obtained ship in \path{reproduce/figures/expected/}.

%%%%%%%%%%%%%%%%%%%%%%%%%%%%%%%%%%%%%%%%%%%%%%%%%%%%%%%%%%%%%%%%%%%%%
\subsection{Notes}

The cuBLAS and cuDNN inputs are library modules containing multiple
kernels. The mapping from each module to the kernels it contains is
recorded in the corpus manifest
(\path{reproduce/sm90_manifest.json}).

%%%%%%%%%%%%%%%%%%%%%%%%%%%%%%%%%%%%%%%%%%%%%%%%%%%%%%%%%%%%%%%%%%%%%
\subsection{Methodology}

The artifact follows the
\href{https://www.acm.org/publications/policies/artifact-review-and-badging-current}
{ACM Artifact Review and Badging guidelines} and the
\href{https://ctuning.org/ae}
{Tuning Artifact Evaluation Methodology}.

\newpage

\bibliographystyle{IEEEtran}
\bibliography{paper}

\end{document}